\documentclass[apj,twocolumn]{emulateapj}
\usepackage{natbib}
\usepackage{url}

\newcommand{\mjysr}{MJy sr$^{-1}$}
\newcommand{\sfmc}{S$^4$MC}

\shorttitle{SMC PAH Spectroscopy}
\shortauthors{Sandstrom et al.}

\begin{document}


\submitted{Submitted to ApJ}

\title{The Spitzer Spectroscopic Survey of the Small Magellanic Cloud
(S$^4$MC): Probing the Physical State of Polycyclic Aromatic
Hydrocarbons in a Low-Metallicity Environment}

\author{Karin M. Sandstrom\altaffilmark{1,2}, 
        Alberto D. Bolatto\altaffilmark{3},
        Caroline Bot\altaffilmark{4,5},
        B. T. Draine\altaffilmark{6},
        James G. Ingalls\altaffilmark{7},
        Frank P. Israel\altaffilmark{8},
        James M. Jackson\altaffilmark{9},
        Adam K. Leroy\altaffilmark{10},
        Aigen Li\altaffilmark{11},
        M\'{o}nica Rubio\altaffilmark{12},
        Joshua D. Simon\altaffilmark{13},
        J. D. T. Smith\altaffilmark{14},
        Sne\v{z}ana Stanimirovi\'{c}\altaffilmark{15},
        A. G. G. M. Tielens\altaffilmark{8},
        Jacco Th. van Loon\altaffilmark{16}}

\affil{$^1$Max Planck Institut f\"{u}r Astronomie, K\"{o}nigstuhl 17, D-69117, Heidelberg Germany}
\affil{$^2$Astronomy Department, 601 Campbell Hall, University of
  California, Berkeley, CA 94720, USA} 
\affil{$^3$Department of Astronomy and Laboratory for Millimeter-wave
  Astronomy, University of Maryland, College Park, MD 20742, USA}
\affil{$^4$Universit\'{e} de Strasbourg, Observatoire Astronomique de Strasbourg}
\affil{$^5$CNRS, Observatoire Astronomique de Strasbourg, UMR7550, F-67000 Strasbourg, France}
\affil{$^6$Department of Astrophysical Sciences, Princeton University, Princeton, NJ 08544, USA}
\affil{$^7$Spitzer Science Center, California Institute of Technology, Pasadena, CA 91125, USA}
\affil{$^8$Sterrewacht Leiden, Leiden University, PO Box 9513, 2300 RA Leiden, The Netherlands}
\affil{$^{9}$Institute for Astrophysical Research, Boston University, Boston, MA 02215, USA}
\affil{$^{10}$National Radio Astronomy Observatory, 520 Edgemont Road, Charlottesville, VA 22903, USA; Hubble Fellow}
\affil{$^{11}$Department of Physics and Astronomy, University of Missouri, Columbia, MO 65213, USA}
\affil{$^{12}$Departamento de Astronomia, Universidad de Chile, Casilla 36-D, Santiago, Chile}
\affil{$^{13}$Observatories of the Carnegie Institution of Washington, 813 Santa Barbara Street, Pasadena, CA 91101, USA}
\affil{$^{14}$Ritter Astrophysical Research Center, University of Toledo, Toledo, OH 43603, USA}
\affil{$^{15}$Department of Astronomy, University of Wisconsin, Madison, 475 North Charter Street, Madison, WI 53703, USA}
\affil{$^{16}$Astrophysics Group, Lennard-Jones Laboratories, Keele University, Staffordshire ST5 5BG, UK}

\email{sandstrom@mpia.de}


\begin{abstract}

We present results of mid-infrared spectroscopic mapping observations of
six star-forming regions in the Small Magellanic Cloud from the Spitzer
Spectroscopic Survey  of the SMC (S$^4$MC).  We detect the mid-IR
emission from polycyclic aromatic hydrocarbons (PAHs) in all of the mapped
regions, greatly increasing the range of environments where PAHs have
been spectroscopically detected in the SMC.  We investigate the
variations of the mid-IR bands in each region and compare our results to
studies of the PAH bands in the SINGS sample and in a sample of
low-metallicity starburst galaxies.  PAH emission in the SMC is
characterized by low ratios of the 6$-$9 \micron\ features relative to
the 11.3 \micron\ feature and weak 8.6 and 17.0 \micron\ features.
Interpreting these band ratios in the light of laboratory and
theoretical studies, we find that PAHs in the SMC tend to be smaller and
less ionized than those in higher metallicity galaxies. Based on studies
of PAH destruction, we argue that a size distribution shifted towards
smaller PAHs cannot be the result of processing in the interstellar
medium, but instead reflects differences in the formation of PAHs at low
metallicity. Finally, we discuss the implications of our observations
for our understanding of the PAH life-cycle in low-metallicity
galaxies---namely that the observed deficit of PAHs may be a consequence
of PAHs forming with smaller average sizes and therefore being more
susceptible to destruction under typical interstellar medium conditions.

\end{abstract}



\section{Introduction}\label{intro}

The mid-infrared emission from normal star-forming galaxies is
characterized by a  series of bright emission bands at approximately
3.3, 6.2, 7.7, 8.6 and 11.3 \micron,  with weaker features at
surrounding wavelengths.  Laboratory work has shown a close
correspondence between the interstellar emission features and the
vibrational modes of the carbon-carbon (C-C)  and carbon-hydrogen (C-H)
bonds in polycyclic aromatic hydrocarbons
\citep[PAHs;][]{leger84,allamandola85,allamandola89}. Although there is
not a definitive laboratory identification of the mid-IR band carrier or
population of carriers, it is considered very  likely that these
features arise from the vibrational de-excitation of PAHs with a range
of sizes after absorbing UV or optical photons \citep[for a recent review,
see][]{tielens08}. 

The widespread  observation of the emission bands in the Milky Way and
elsewhere suggests that PAHs  are an abundant and energetically
important  component of interstellar dust \citep[emitting $\sim 10$\% of
the total infrared emission from a galaxy;][]{helou00,smith07a}. In
addition,  PAHs play a number of important roles in the interstellar
medium (ISM).  They are a major source of  photoelectric heating
\citep{bakes94,wolfire95,hollenbach99,helou01} and participate in
chemical reaction networks \citep{bakes98,wolfire08,wakelam08} in a
variety of ISM  phases.  Because of these crucial roles, we would like
to understand how and why the characteristics of PAHs depend on galaxy
properties  such as metallicity and star-formation history and what
specific factors control the abundance, size distribution and physical
state (ionization, hydrogenation, compactness, symmetry, etc.) of PAHs
in the ISM.

The physical state of a population of PAHs is reflected in the relative
intensities, profile shape and central wavelength of the mid-IR emission
bands.  Of the major PAH bands, laboratory spectroscopy has suggested
that those at 3.3, 8.6 and 11$-$14 \micron\ arise from C-H stretching
and bending modes while the 6.2 and 7.7 complexes are from C-C
stretching.  Longer wavelength features like the 17.0 \micron\ complex
are most likely related to more complex ring bending modes of the
molecule \citep{moutou00}. The relative strengths of these bands for a
given population of PAHs will depend on their size distribution and
physical state allowing us diagnostics of the PAH population from mid-IR
spectroscopy.

For instance, laboratory and theoretical investigations have
consistently shown that the C-H out-of-plane bending modes between
11$-$15 \micron\ in ionized PAHs are weak, while they are strong in
neutral PAHs.  The opposite is true of the C-C stretch and C-H in-plane
bending modes between $6-9$ \micron\ \citep{szczepanski93,hudgins94}, so
the ratios of these two sets of features should trace PAH ionization to
some degree.  Recent work by \citet{galliano08a} argued that the
variations of the 6.2, 7.7, 8.6 and 11.3 \micron\ features within and
between galaxies are consistent with being controlled by ionization.
Specifically, they find that the ratios of 6.2, 7.7 and 8.6 to the 11.3
\micron\ bands are correlated over an order-of-magnitude, while ratios
such as 7.7/6.2 and 8.6/6.2 stay roughly constant as a function of the
7.7/11.3 ratio.  

The ratios of short-to-long wavelength PAH bands, however, are
also expected to depend on the PAH size distribution \citep{draine01},
such that small PAHs emit more strongly in shorter wavelength features.
In addition to this general trend, laboratory and theoretical
studies have suggested that some specific features originate from large
PAHs.  \citet{peeters04b} find that large PAHs are required to
reproduce the general characteristics of the 17.0 \micron\ complex,
although the charge state of the carrier is not well determined. Recent
work by \citet{bauschlicher08,bauschlicher09} has suggested that the
strength of the 8.6 \micron\ band also is an indicator of the
contribution of large PAHs to the mid-IR spectrum.  They find that in
their theoretical investigations of the spectrum of large compact and
irregular PAHs, only large, charged, compact PAHs have a feature at 8.6
\micron\ that can reproduce the interstellar band.  

Finally, as reviewed in detail by \citet{hony01} and \citet{bauschlicher09}, the
relative strengths of the 11$-$15 \micron\ C-H out-of-plane bending
modes provide information about the structure of PAHs.  The type of
structure adjacent to the vibrating C-H bond determines the wavelength
of the mode.  If a given aromatic carbon ring on the edge of a PAH has
only one attached hydrogen (out of four possible for a PAH that consists
of more than one aromatic ring), while the rest of the bonds are
connected to other aromatic rings, it is considered a ``solo'' hydrogen.
If two of the possible sites contain C-H bonds, this is considered a
``duo'', and the ``trio'' and ``quartet'' are similarly defined.  In
structural terms, straight edges produce ``solo'' hydrogens while
corners produce ``duo'' hydrogens.  Irregularities like bays and linear
extensions produced ``trio'' and ``quartet'' hydrogens.  These different
modes have different wavelengths \citep[dependent on the size of the
molecule;][]{bauschlicher09}. Thus, observations of the strengths of the
11$-$15 \micron\ features can provide information about the PAH
structure.

Using the knowledge of PAH bands gained from these laboratory and
theoretical studies, it is possible to diagnose the physical state of
PAHs in extragalactic environments observed with ISO and Spitzer.  In
particular, evidence for changes in PAH properties with radiation field
intensity and metallicity have been seen in a variety of studies:  

\begin{itemize}

\item \citet{galliano08a} studied the variation of the ratios of the
      6.2, 7.7, 8.6 and 11.3 bands in a diverse sample of galaxies and
      individual star-forming regions and found variations in the 6$-$9
      vs 11.3 \micron\ features correlated with the local UV field
      strength---a trend they interpreted as arising mainly from changes
      in the PAH ionization state. 

\item \citet{smith07a} have found that PAH emission from AGN-hosting
      galaxy nuclei shows very low 7.7/11.3 ratios \citep[see
      also][]{odowd09,diamond-stanic10,wu10}.  Some studies have also
      found that the spectra of these sources also show average or even
      high ratios of the 17.0 feature to the 7.7 and 11.3 bands
      \citep{smith07a,odowd09}.  These trends have been interpreted as
      the preferential destruction of small PAHs by the hard radiation
      field produced by the AGN, leading to decreased emission in the
      short wavelength bands relative to the long wavelength bands.  

\item \citet{hunt10} studied the PAH features in a sample of blue
      compact dwarfs (BCDs) with a range of metallicities and found
      evidence for increased strengths of the 8.6 and 11.3 features
      relative to the other PAH bands in these intensely star-forming
      galaxies.  They interpreted these trends as evidence for
      destruction of small PAHs in the harder, more intense radiation
      fields of these galaxies citing recent theoretical work by
      \citet{bauschlicher08} which suggests that the 8.6 \micron\
      feature arises predominantly from large PAHs.

\item \citet{smith07a} measured the integrated PAH band strengths in the
      central regions of galaxies in the SINGS sample and found
      indications that the 17.0 \micron\ complex is weak compared to the
      other PAH bands at low-metallicity, although their sample contains
      few low-metallicity galaxies.   

\item Resolved studies of PAH emission in the Magellanic Clouds have
      also shown variations in the band ratios.  For instance,
      \citet{reach00} detected PAH emission from the quiescent molecular
      clouds SMC B1 \#1 with a very low 7.7/11.3 ratio.  \citet{li02}
      found that they could not reproduce the SMC B1 ratio using their
      Milky Way PAH model.  
      
\end{itemize}

In addition to these changes in the PAH physical state, observations
with ISO and Spitzer have demonstrated that the fraction of PAHs
relative to the total amount of dust decreases in low-metallicity
environments
\citep{madden00,engelbracht05,madden06,wu06,draine07b,gordon08,hunt10,lebouteiller11}.
A number of explanations for the PAH deficit in low-metallicity galaxies
have been proposed including a time lag between the formation of PAHs in
AGB star atmospheres relative to SNe-produced dust \citep{galliano08b},
destruction of PAHs by supernova shocks \citep{ohalloran06} and
destruction of PAHs by harder and/or more intense UV fields
\citep{madden06,gordon08}.  Recent work by \citet{sandstrom10} (Paper I) on the
fraction of PAHs in the Small Magellanic Cloud has shown a very low PAH
fraction in the diffuse ISM of this low-metallicity galaxy and a PAH
fraction in dense gas within a factor $\sim 2$ of the Milky Way.  They
argue that PAHs are destroyed in the diffuse ISM of the SMC and may be
forming in dense regions.  Thus, in addition to variations in the
destruction of PAHs, it is possible that as a function of metallicity
the balance between AGB-produced PAHs and PAHs that form in the ISM may
change.  

These proposed variations in the life-cycle of PAHs in low-metallicity
galaxies will be reflected in changes to the physical state of the PAHs
themselves --- their size distribution, hydrogenation, ionization state,
chemical substitutions and/or functional groups and structural
characteristics.  Enhanced destruction of PAHs by UV fields or shocks
should change the PAH size distribution, since smaller molecules are
easier to photodissociate \citep{allain96a,allain96b,lepage01}.  If the
dominant formation mechanism of PAHs changes as a function of
metallicity it may be reflected in the size distribution and
irregularity of the molecules.  Thus, by observing the physical state of
PAHs in a low-metallicity environment we hope to understand what drives
the observed differences in the PAH life-cycle.

To that end, we present the results of a mid-IR spectroscopic survey of
PAH emission in the Small Magellanic Cloud (SMC).  At a distance of 61
kpc \citep{hilditch05} and with a metallicity of $12 +$ log(O/H) $\sim
8.0$ \citep{kurt98}, the SMC presents the ideal target to understand the
properties of PAHs in low-metallicity systems at high spatial
resolution. The SMC represents an intermediate case between Galactic
sources where variations can be studied within objects
\citep{hony01,peeters02,rapacioli05,berne07} and other galaxies where
only galaxy-wide results are obtained.  The Spitzer Spectroscopic Survey
of the SMC (\sfmc) consists of low-resolution spectral mapping
observations using the Infrared Spectrograph (IRS) on Spitzer over six
of the major star-forming regions of the SMC. In Paper I, we presented
the results of SED and spectral fitting using the \sfmc\ data to
determine the PAH fraction across the galaxy.  In this paper we turn our
attention to using the band ratios to learn about the physical state of
PAHs.  

In Section~\ref{sec:s4mc} we describe the \sfmc\ observations and data
reduction.  Section~\ref{sec:fits} covers the spectral fitting.
In Section~\ref{sec:results} we describe the variations in the band
ratios in the six star-forming regions in our sample and finally in
Section~\ref{sec:interp} we discuss what the variations tell us about
the physical state of SMC PAHs and in Section~\ref{sec:lc}, the
implications of our results for understanding the PAH deficiency in
low-metallicity galaxies.

\section{Spitzer Spectroscopic Survey of the Small Magellanic Cloud
(\sfmc)}\label{sec:s4mc}

\begin{figure*}
\centering
\epsscale{1.0}
\includegraphics[width=7in]{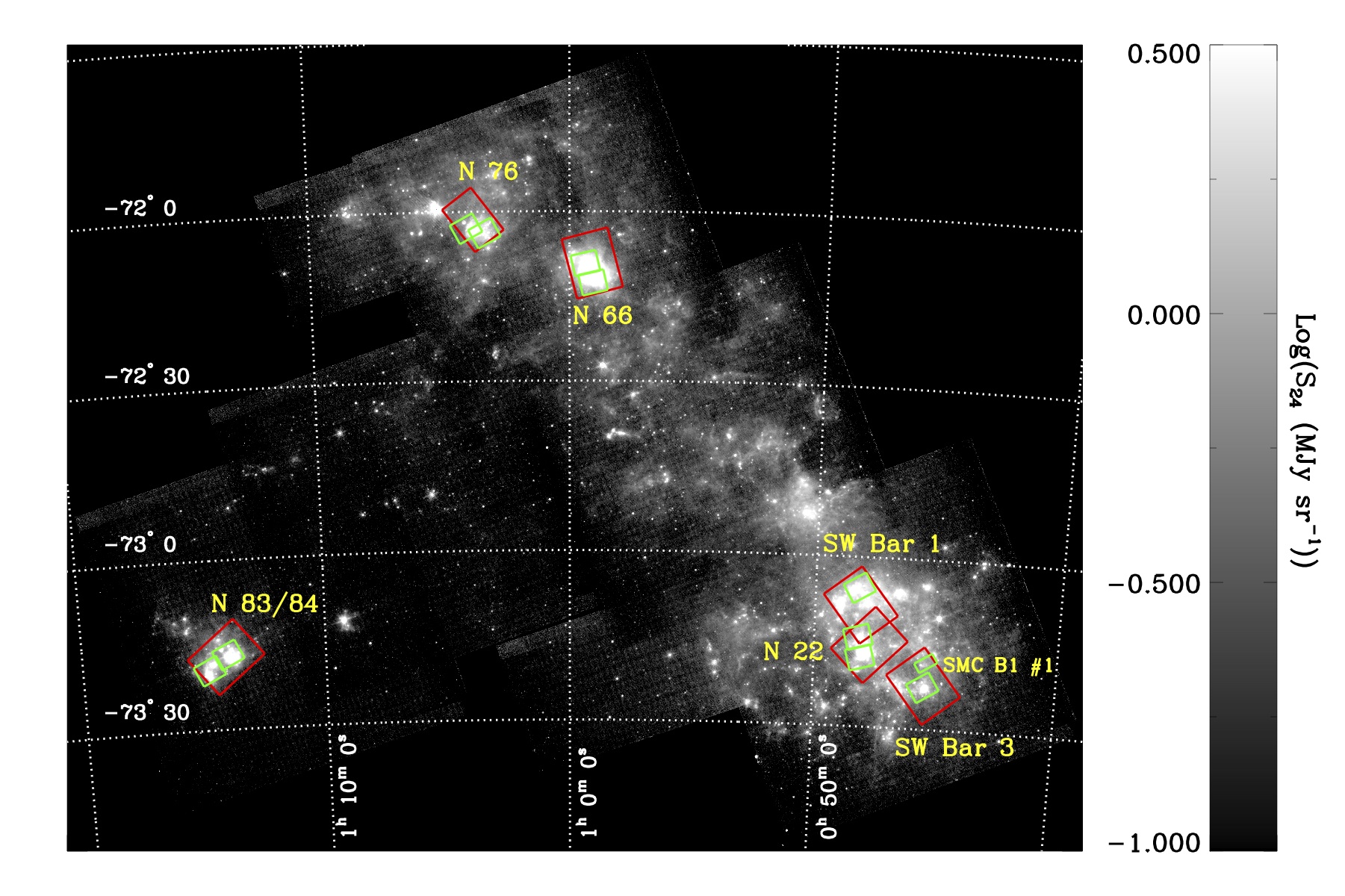}
\caption{The coverage of the S$^4$MC survey overlayed on the MIPS 24
\micron\ map from S$^3$MC.  The color scale is logarithmic, with the
stretch illustrated in the colorbar.  The red boxes show the coverage
of the LL1 order maps (the LL2 maps are shifted by $\sim 3\arcmin$)
and the green boxes show the coverage of the SL1 order maps (the SL2
maps are shifted by $\sim 1\arcmin$).  We also identify the various
regions of the galaxy by the names we will refer to in the remainder
of this paper.}
\label{fig:coverage}
\end{figure*}

\begin{figure*}
\centering
\epsscale{2.4}
\plottwo{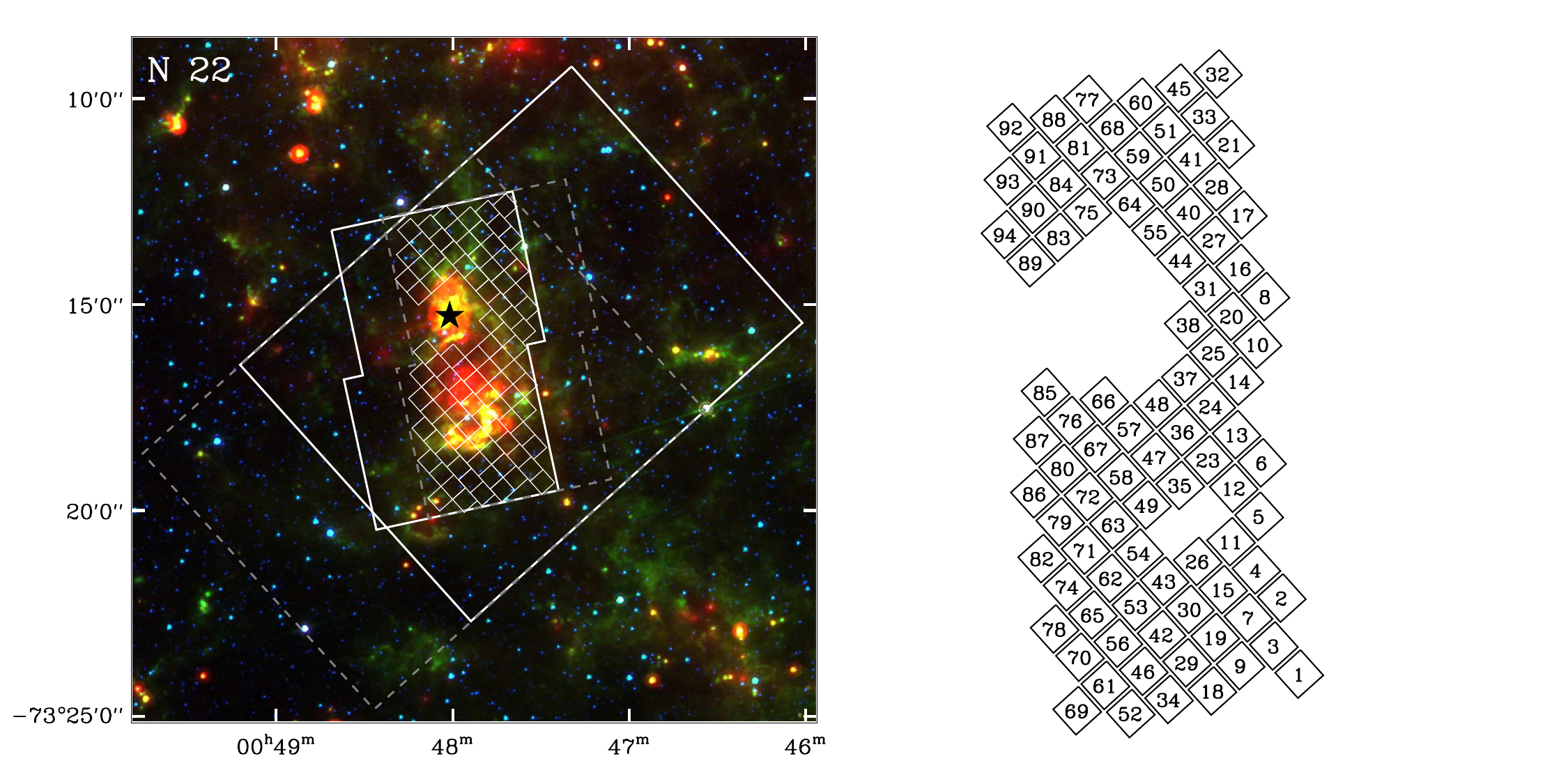}{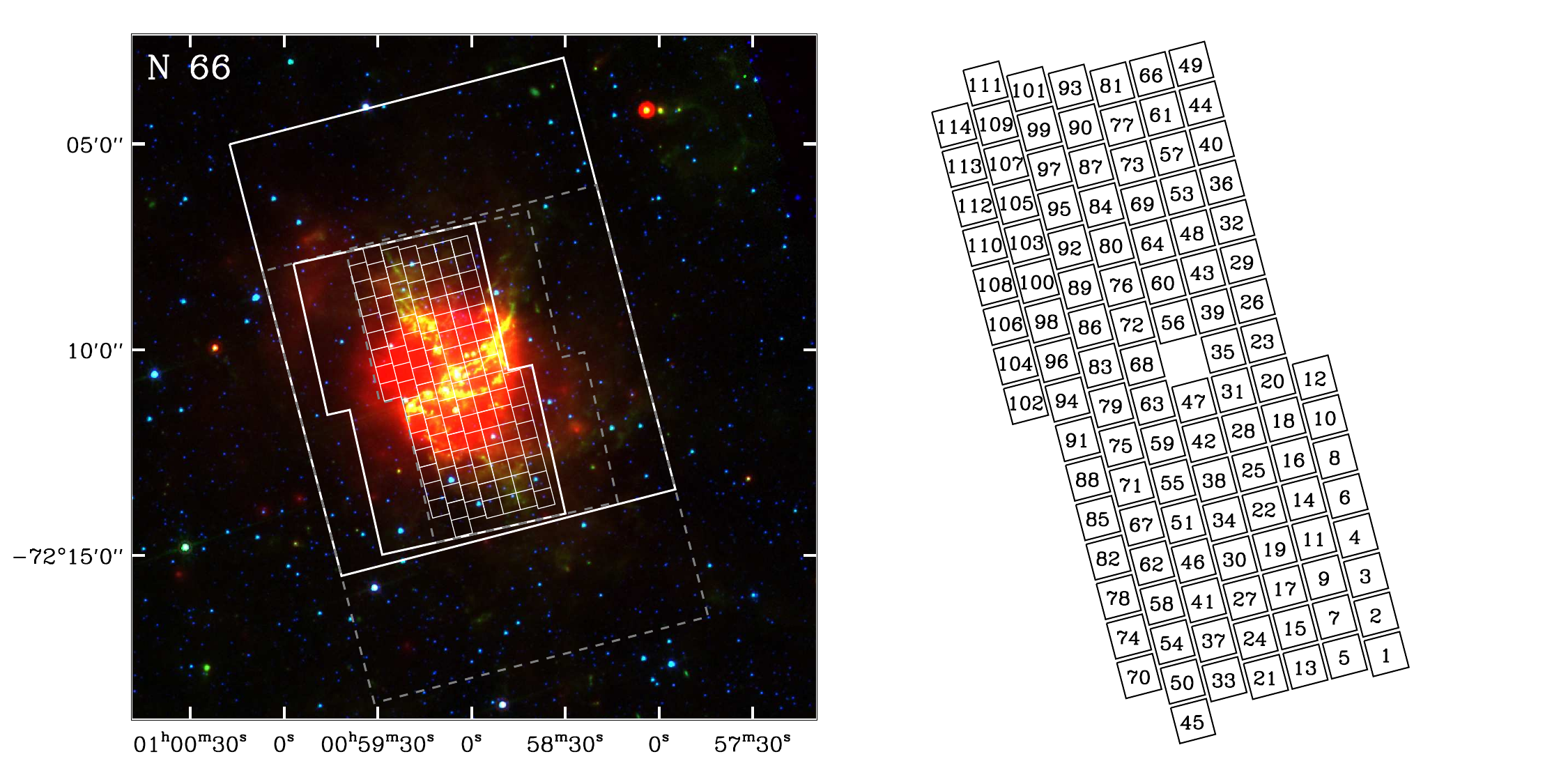}
\caption{Coverage of the S$^4$MC observations in the N 22 and N 66
regions.  The three-color images are constructed from S$^3$MC photometry
in the IRAC 3.6 (blue), IRAC 8.0 (green) and MIPS 24 (red) bands.  The
color table is the same in Figures~\ref{fig:coverage1},
~\ref{fig:coverage2} and ~\ref{fig:coverage3}.  The coverage of the LL1
and SL1 orders is shown with a solid white line while the LL2 and SL2
coverage is shown with a gray dashed line.  For the following analysis
we are limited to regions with coverage in both SL bands.  The white
boxes illustrate locations where we have extracted spectra from the cube
for this analysis (see the text for details) and the right panel shows
all of the extracted regions along with a number identifying them.
There is a very bright point source in the N 22 region which is
saturated in the 24 \micron\ MIPS observation, we show its location with
a black star. Because of the artifacts associated with bright point
sources we are forced to omit some regions from our analysis.}
\label{fig:coverage1}
\end{figure*}

\begin{figure*}
\centering
\epsscale{2.4}
\plottwo{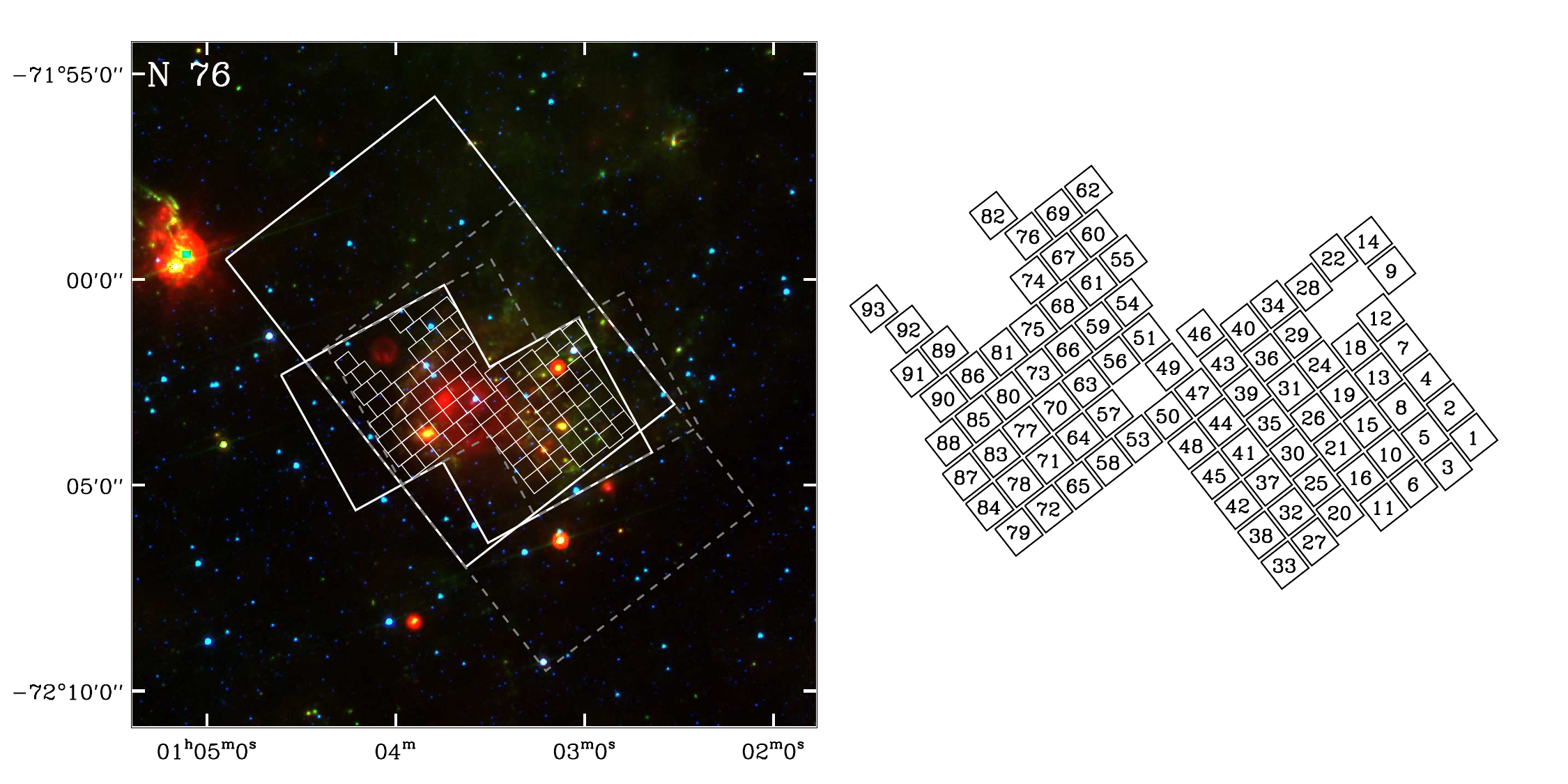}{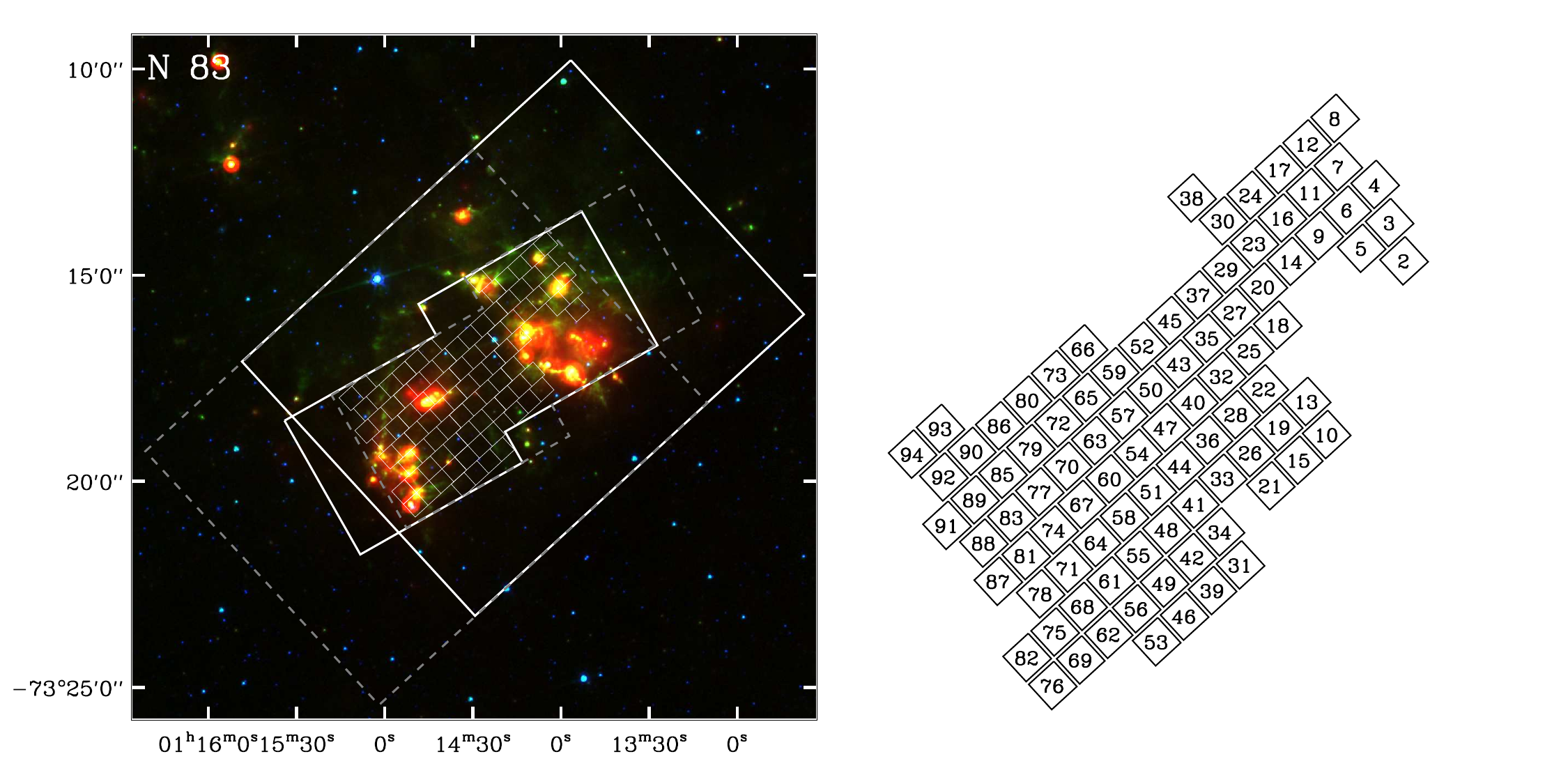}
\caption{Coverage of the S$^4$MC observations in N 76 and N 83.  The
images and color table are the same as in Figure~\ref{fig:coverage1}. In
N 76, we avoid the region dominated by the supernova remnant
1E\,0102.2$-$7219 \citep[visible as a ring-like feature in the
24\micron\ band;][presented an analysis of the IRS observations for this
region]{sandstrom09}. In N 83, most of the area of the northern SL cubes
are affected by artifacts due to the saturation of the peak-up array due
to a very bright point source in the region.  We avoid those areas in
selecting the extraction regions.} 
\label{fig:coverage2}
\end{figure*}

\begin{figure*}
\centering
\epsscale{2.4}
\plottwo{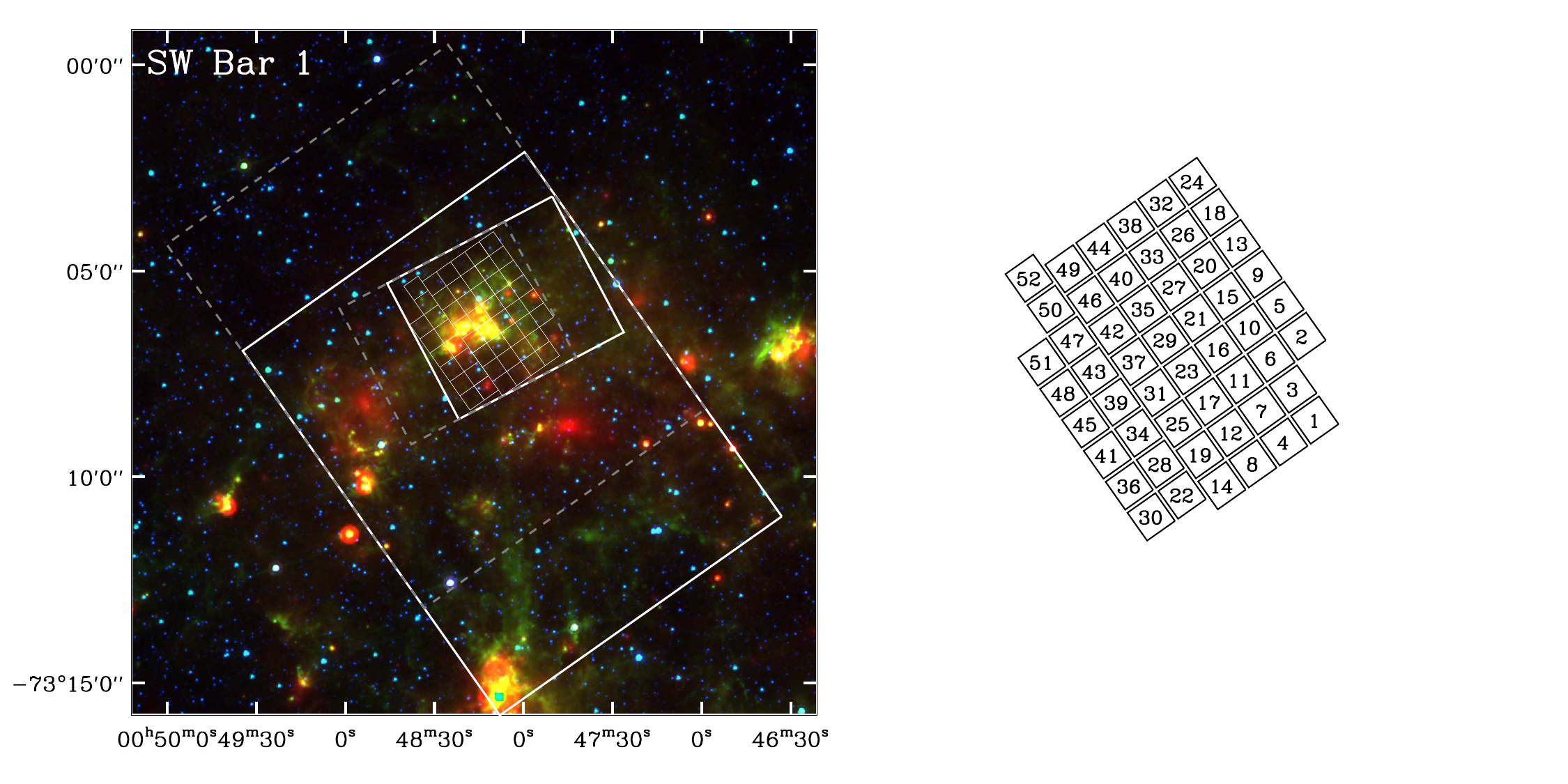}{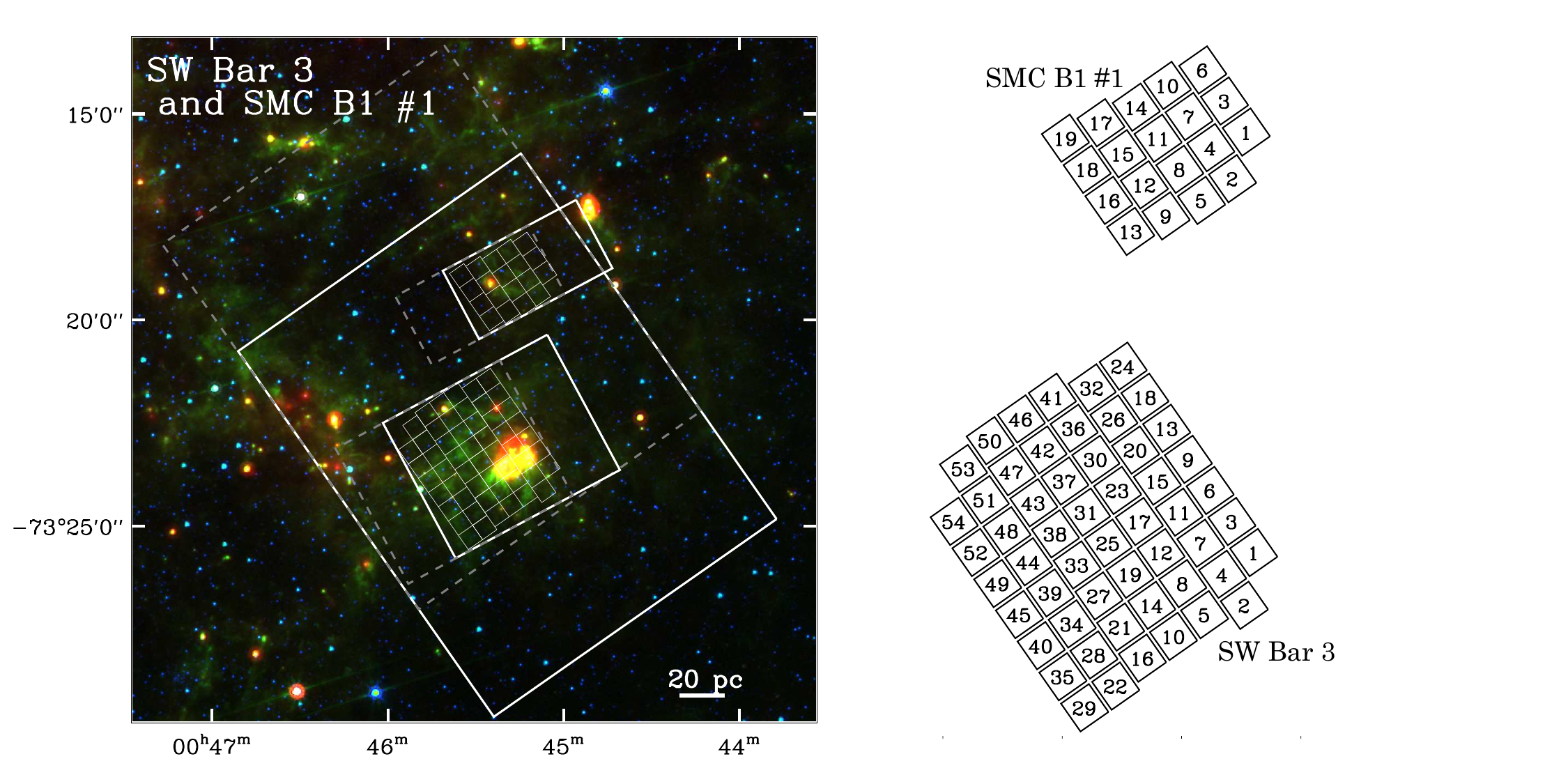}
\caption{Coverage of the S$^4$MC observations in the SW Bar 1 and 3
regions.  The images and color table are the same as in
Figure~\ref{fig:coverage1}. The coverage of the SMC B1 spectral map is
shown in the SW Bar 3 region.}
\label{fig:coverage3}
\end{figure*}

We mapped six star-forming regions in the SMC using the low-resolution
orders of the InfraRed Spectrograph (IRS) on Spitzer as part of the
\sfmc\ project \citep[GO 30491---now combined with a larger mosaic of
the galaxy as part of SAGE-SMC;][Gordon et al. 2011,
submitted]{bolatto07}.  Figure~\ref{fig:coverage} shows the coverage of
our spectral maps overlaid on the S$^3$MC 24 \micron\ mosaic
\citep{bolatto07}.  Figures~\ref{fig:coverage1},~\ref{fig:coverage2}
and~\ref{fig:coverage3} show three-color images of the six regions,
overlaid with the coverage of all of the orders.  These regions are all
displayed with the same color scale.  In the following analysis, we are
limited to the regions with coverage in both of the SL orders because of
our focus on the PAH emission bands.  We list the details of the
observations in Table~\ref{tab:s4mc}.

The regions we have mapped were selected to cover a range of properties
and therefore represent a diverse sample of SMC sources.  The N 66 field
covers the well studied giant H II region NGC 346, which hosts 33 O
stars \citep{massey89}.  ISO observations of N 66 by \citet{contursi00}
show radiation field intensity upwards of $10^5$ times the solar
neighborhood radiation field in some of the brightest mid-IR peaks.  We
have also mapped the quiescent molecular region SMC B1 \#1, found in the
SW Bar \citep{rubio93}.  This was the first location where PAH emission
was detected spectroscopically in the SMC by \citet{reach00}.  Also in
the SW Bar region, we map three other fields: N22, a compact H II region
surrounded by molecular gas; SW Bar 1, a region of active, but more
evolved, star-formation (including the N 27 nebula), with bright 8
\micron\ and H$\alpha$ emission; and SW Bar 3, which covers regions of
molecular gas and recent star-formation.  In the Wing of the SMC, we map
the N 83/N 84 complex, which covers a region of high UV field intensity
with abundant molecular gas \citep{leroy09}.  Finally, in the N 76
region our map covers a young supernova remnant 1E\,0102.2$-$7219
\citep{stanimirovic05,sandstrom09}, a Wolf-Rayet nebula and a few small
regions of molecular gas on the outskirts of the nebula.

\begin{deluxetable*}{lcccc}
\tablewidth{0pt}
\tabletypesize{\scriptsize}
\tablecolumns{5}
\tablecaption{S$^4$MC Observations}
\tablehead{\multicolumn{1}{l}{Region} &
\multicolumn{1}{c}{R. A. (J2000)} &
\multicolumn{1}{c}{Dec. (J2000)} &
\multicolumn{1}{c}{Observation Date} &
\multicolumn{1}{c}{AOR Number}} 
\startdata
\cutinhead{LL Observations}
SW Bar 3 & $0^{\rm h}45^{\rm m}23^{\rm s}$ & $-73^\circ22\arcmin00\arcsec$ & 11 Jun 2007  & 18264832 \\
N 22     & $0^{\rm h}47^{\rm m}42^{\rm s}$ & $-73^\circ15\arcmin06\arcsec$ & 09 Sep 2006  & 18261504 \\
SW Bar 1 & $0^{\rm h}48^{\rm m}08^{\rm s}$ & $-73^\circ08\arcmin24\arcsec$ & 11 Jun 2007  & 18264576 \\
N 66     & $0^{\rm h}59^{\rm m}09^{\rm s}$ & $-72^\circ09\arcmin14\arcsec$ & 15 Nov 2006  & 18262272 \\
N 76     & $1^{\rm h}03^{\rm m}41^{\rm s}$ & $-72^\circ01\arcmin30\arcsec$ & 12 Dec 2006  & 18263040 \\
N 83/84  & $1^{\rm h}14^{\rm m}05^{\rm s}$ & $-73^\circ17\arcmin00\arcsec$ & 10 Sep 2006  & 18263808 \\
\cutinhead{SL Observations}
SW Bar 3   & $0^{\rm h}45^{\rm m}26^{\rm s}$ & $-73^\circ22\arcmin30\arcsec$ & 10 Jun 2007 & 18265344 \\ 
SMC B1 \#1 & $0^{\rm h}45^{\rm m}21^{\rm s}$ & $-73^\circ20\arcmin10\arcsec$ & 10 Jun 2007 & 18261248 \\ 
SW Bar 1   & $0^{\rm h}48^{\rm m}13^{\rm s}$ & $-73^\circ05\arcmin20\arcsec$ & 10 Jun 2007 & 18265088 \\ 
N 22 North & $0^{\rm h}48^{\rm m}11^{\rm s}$ & $-73^\circ13\arcmin56\arcsec$ & 20 Nov 2006 & 18262016 \\ 
N 22 South & $0^{\rm h}48^{\rm m}04^{\rm s}$ & $-73^\circ17\arcmin50\arcsec$ & 20 Nov 2006 & 18261760 \\ 
N 66 North & $0^{\rm h}59^{\rm m}22^{\rm s}$ & $-72^\circ09\arcmin07\arcsec$ & 21 Nov 2006 & 18262784 \\ 
N 66 South & $0^{\rm h}59^{\rm m}05^{\rm s}$ & $-72^\circ12\arcmin50\arcsec$ & 20 Nov 2006 & 18262528 \\ 
N 76 1     & $1^{\rm h}03^{\rm m}11^{\rm s}$ & $-72^\circ04\arcmin01\arcsec$ & 09 Dec 2006 & 18263296 \\ 
N 76 2     & $1^{\rm h}04^{\rm m}00^{\rm s}$ & $-72^\circ02\arcmin50\arcsec$ & 09 Dec 2006 & 18263552 \\ 
N 83       & $1^{\rm h}14^{\rm m}00^{\rm s}$ & $-73^\circ16\arcmin40\arcsec$ & 10 Dec 2006 & 18264064 \\ 
N 84       & $1^{\rm h}14^{\rm m}50^{\rm s}$ & $-73^\circ19\arcmin20\arcsec$ & 10 Dec 2006 & 18264320
\enddata
\label{tab:s4mc}
\end{deluxetable*}

\subsection{Mapping Strategy and Spectral Cube Construction}

The maps were constructed by stepping the IRS slit in a grid of
positions parallel and perpendicular to the slit.  The maps are fully
sampled by stepping by half of a slit width (steps of 1.85\arcsec\ and
5.08\arcsec\ for SL and LL, respectively) perpendicular to the slit. The
LL maps are composed of a grid of 98 perpendicular and 7 parallel steps,
with half slit length steps parallel to the slit and 14 seconds of
integration time per position (except N 76 which has 75 $\times$ 6 steps
perpendicular and parallel).  The LL spectral maps cover an area of
$493\arcsec \times 474\arcsec$ ($376\arcsec \times 395\arcsec$ for N
76). The SL maps are constructed similarly except that we use full slit
width steps parallel to the slit to increase the mapped area at the
expense of some pixel redundancy.  All of the SL maps, aside from the
map of SMC B1, have 120 perpendicular by 5 parallel pointings
($220\arcsec \times 208\arcsec$) with integration times of 14 seconds
per position.  We have made a deeper map of SMC B1, the location where
PAH emission was first detected from the SMC \citep{reach00} using 60 by
4 slit positions ($109\arcsec \times 156\arcsec$) with 60 second
integration times.  The spectral coverage of the low-resolution orders
extends between 5.2 and 38.5 \micron\ with spectral resolving power
ranging between $\sim 60-120$.  

Each mapping observation was preceded or followed closely in time by an
observation of an ``off'' position at RA 1$^{\rm h}$9$^{\rm
m}$40$^\mathrm{s}$ Dec $-$73$^\circ$31$\arcmin$30$\arcsec$ (J2000).
This position was chosen to be free from SMC emission at the level of
our MIPS and IRAC mosaics from S$^3$MC. The ``off'' position was
observed with the same ramp times as its corresponding mapping
observations using ``staring mode''.  There were four nod positions in
each ``off'' observation, with 9 repetitions per nod.  In the course of
processing the maps, the outlier-clipped mean of the 36 two-dimensional
spectral images of the ``off'' position was subtracted from each of the
mapping BCDs.  This removes the zeroth-order foreground emission from
the zodiacal light and Milky Way cirrus as well as mitigating the
effects of time-variable rogue pixels in the low-resolution arrays.

We have used the BCDs from pipeline version S15.3 and S16.1 to create
our final cubes.  There are no significant differences between these two
pipeline versions for our purposes. The cubes were assembled using the
Cubism software package (v1.7) distributed by the SSC \citep{smith07b}.
Cubism takes the pipeline processed BCDs as input and uses
polygon-clipping based re-projection to transform the two dimensional
spectral images into a three dimensional spectral cube.  Cubism also
propagates the uncertainties associated with the data collection events,
known as ramps.  We propagate these uncertainties throughout the
following analysis, but note that they represent only the statistical
uncertainties of each pixel and do not take into account the more
important systematic errors.  In assembling the cubes in Cubism we apply
a ``slit-loss correction function'' which is essentially an extended
source calibration factor for the IRS \citep{smith07b}.

We remove bad pixels in the cubes using the tools provided in Cubism.
These bad pixels are identified by the stripes they produce in the
spectral maps, resulting from the movement of the bad pixel through the
grid of positions.  The number of bad pixels increases towards the
longest wavelengths in the cube, effectively making the spectrum past
$\sim 35$ \micron\ have much lower signal-to-noise.  Regarding bad pixel
flagging, we note that in cases where the core of the PSF of a point
source falls on a bad pixel in the detector, there can be difficulties
in reconstructing the emission profile in Cubism. This issue is
exacerbated by two problems: the spatial under-sampling of the IRS, and
the ``full-width'' steps used to construct the map, which reduces the
pixel redundancy. Because we are interested in the diffuse emission, we
have performed extensive bad pixel subtraction in order to get the
cleanest map of the region without regard to the effects on point
sources.  For that reason, in the vicinity of bright point sources there
can be artifacts that interfere with the determination of the PAH
emission band strengths in the spectra of point sources .  In the
following, we avoid these regions in the cubes, particularly in the core
of N 66 and in the vicinity of the bright point sources in N 22 and N
83.  In N 83, the brightest point source presents another difficulty in
that it saturates the peak-up array, leading to uncorrected droop
effects over a large fraction of the map.  In our analysis, we have
avoided regions affected by this artifact as well.

\subsection{Foreground Subtraction}

The ``off'' position observations that were performed close in time to
each mapping BCD allow us to subtract the majority of the zodiacal and
Milky Way foregrounds from our maps.  There are, however, gradients in
both foregrounds between the ``off'' location and the map location.  To
remove these additional gradients we use the zodiacal light spectrum
predicted by the Spitzer Observation Planning Tool (SPOT) at the times
of each observation to calculate the difference in the zodiacal
foreground between the ``off'' location and the map location versus
wavelength.  SPOT uses the zodiacal light model of \citet{kelsall98}
from DIRBE observations.  The modifications to this model for predicting
the zodiacal light at Spitzer wavelengths are described in the
documentation from the Spitzer Science
Center(SSC)\footnote{\url{http://ssc.spitzer.caltech.edu/documents/background/ \\ bgdoc\_release.html}}.
To determine the gradient in the Milky Way foreground we assume a
correlation between the cirrus dust emission and the column density of
neutral hydrogen \citep[as in][]{boulanger96}.  We use the
\citet{draine07a} emissivity of Milky Way dust and a map of H I from a
combined ATCA/Parkes survey (E. Muller, private communication) to
measure the Milky Way foreground gradient as in Paper I.

\subsection{``Dark Settle'' Related Artifacts in the \sfmc\ Spectral
Mapping}\label{sec:artifact}

In performing the analysis of the \sfmc\ data we discovered that both
the SL and LL observations are affected by an artifact related to a
time- and position variable residual background level on the detector.
This artifact is a general issue in IRS spectral mapping, but is
particularly important for faint regions like those we study here.  This
issue is very similar to what has been seen in LH observations and
referred to as the ``dark settle''
issue\footnote{\url{http://irsa.ipac.caltech.edu/data/SPITZER/docs/irs/features/}}.
The artifact causes a tilt to the various orders, leading to a mis-match
between the spectral orders LL1-LL2 and SL1-SL2 as well as introducing
spurious curvature to the spectrum.  Because of the faintness of the PAH
emission in many SMC regions, this artifact can have a significant
effect on the flux in the 7.7 \micron\ complex, located at the overlap
of the SL1 and SL2 orders.

Our basic approach to dealing with the artifact is to correct the data
at the BCD level by determining the shape of the residual background
from the inter-order region.  The details of this procedure are
described in the Appendix.  For the SL orders in particular, the
inter-order space is very small, especially since the gap between the
SL3 and SL2 orders is strongly affected by the scattered light from the
peak-up arrays and its correction in the IRS pipeline.  At a basic
level, we do not have information about the background level within the
spectral orders, which is precisely where we would like to remove it.
Therefore, our correction of this artifact is approximate at best and we
proceed in the remainder of the paper to deal with two versions of the
spectral cubes: one with no correction of the ``dark settle'' issue and
one with a correction we determine from the inter-order light.  As
described in the Appendix, the technique we use may overcorrect (i.e. by
removing too much flux from the 7.7 \micron\ feature).  We argue that
the uncorrected and corrected spectra should bracket the range of
possible values for the band strengths.

\subsection{Spectral Extraction and Stitching}

In order to get better signal-to-noise for determining the band ratios
we extract the spectra in $25\arcsec\times 25$\arcsec\ boxes. This
extraction aperture is also large enough to avoid problems with the
variation of the PSF with wavelength (the FWHM of the PSF at 35 \micron\
is $\sim 9$\arcsec). We have tiled the cubes in the region of full
overlap between all orders with these extraction boxes.  The right
panels of Figures~\ref{fig:coverage1},~\ref{fig:coverage2}
and~\ref{fig:coverage3} show the extraction boxes for each region,
excluding those affected by point source artifacts.  We refer to the
individual spectra by the number shown on these diagrams throughout the
paper.

After the correction for foreground gradients and ``dark settle''-type
artifacts, our extracted spectra show offsets in the overlap of the SL1
and LL2 orders that vary from region to region, but stay relatively
constant within a given region.  These offsets are additive and on the
order of a few tenths of  \mjysr.  In correcting these offsets we first
extract photometry in matching apertures from the S$^3$MC 24 \micron\
mosaic \citep[which has been foreground subtracted as described
in][]{sandstrom10} and find the average offset at 24 \micron\ for the
MIPS and synthetic IRS photometry for each region.  We then use this
additive region-based offset to tie the IRS spectrum to the MIPS
photometry---a step which is necessary because we have no prior
information on which order is ``correct'' and the 8 \micron\ photometry
relies on an extended source correction for IRAC and is thus less
reliable than the 24 \micron\ photometry.  Next we determine the offset
between the SL and LL orders in their overlap region ($\sim 14.3-14.7$
\micron) and find another additive region-based correction that we then
add to the SL spectra.

After tying the spectra to the MIPS 24 \micron\ photometry and
correcting the LL/SL offset, we stitch the spectra together by
interpolating the long wavelength end of each order onto the grid of the
overlapping shorter wavelength order and averaging the resulting values
together.  The procedures we employ in correcting offsets between the
orders and stitching the spectra together do not significantly affect
our results.

\subsection{Fitting the PAH Emission Bands}\label{sec:fits}

\begin{figure*}
\centering
\epsscale{1.2}
\plotone{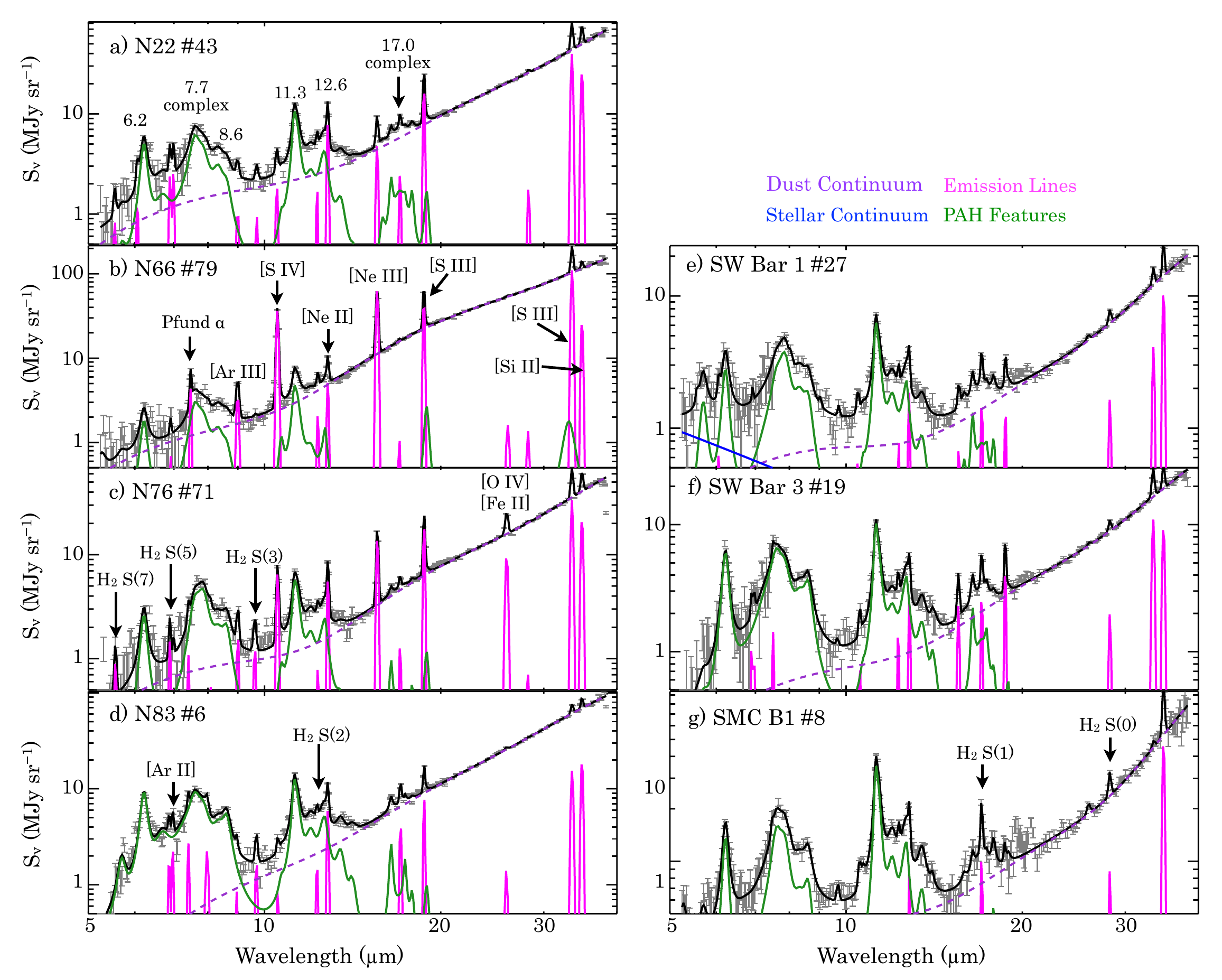}
\caption{These plots show example PAHFIT results for spectra from our
sample.  The spectra are labeled with the region and a number which
corresponds to labels on Figures~\ref{fig:coverage1},
~\ref{fig:coverage2} and~\ref{fig:coverage3}.  Each plot shows the
spectrum and statistical uncertainties in gray, the full best fit model
in black, the PAH contribution to the best fit model in green, the line
emission in magenta, the stellar continuum in blue and the dust
continuum in orange.}
\label{fig:pahfit}
\end{figure*}

We use the PAHFIT spectral fitting routine \citep{smith07a} to measure
the strengths of the PAH emission bands and a variety of emission lines
from our spectra.  Figure~\ref{fig:pahfit} shows the results of fits to
some high signal-to-noise spectra from our dataset.  PAHFIT uses
combinations of Drude profiles to fit the emission bands and Gaussians
to fit the emission lines.  We do not include extinction in our
fits---\citet{lee09} found very low levels of mid-IR extinction in the
SMC as expected given the galaxy's low dust-to-gas ratio.  The effect of
extinction on the band ratios would tend to primarily suppress emission
in the 8.6 and 11.3 \micron\ bands because of the silicate extinction feature at 9.7 \micron.  In the cases where the PAH emission
features form a complex (for instance, the 7.7 \micron\ feature), we
report only the total and not the properties of the individual
components.   We have used the default line list provided in PAHFIT with
the exception of adding in the Pfund-$\alpha$ recombination line of
hydrogen at 7.46 \micron, which we see in the spectra of N 66 and some
of the spectra in N 22.  We perform the fits on the original spectra and
on the ``dark settle'' artifact corrected spectra.

Table~\ref{tab:allmeas} lists the measured integrated intensities for
the major PAH features as well as the lines of [Ne III] at 15.6 \micron\
and [Ne II] at 12.8 \micron\ for each spectrum extracted from the cubes
both with and without the ``dark settle'' correction.

\begin{deluxetable*}{lccrrrrrr}
\tablewidth{0pt}
\tabletypesize{\scriptsize}
\tablecolumns{9}
\tablecaption{Spectral Fit Results}
\tablehead{ \multicolumn{1}{l}{Region} &
\multicolumn{1}{c}{Num} &
\multicolumn{1}{c}{R.A.} &
\multicolumn{1}{c}{Dec.} &
\multicolumn{1}{c}{6.2} &
\multicolumn{1}{c}{7.7} &
\multicolumn{1}{c}{8.3} &
\multicolumn{1}{c}{[Ne II]} &
\multicolumn{1}{c}{[Ne III]} }
\startdata
N 22 &  1 & 11.891834 & $-$73.313254 & 4.237$\pm$0.527 &  5.691$\pm$0.778 &  0.361$\pm$0.371 & 0.213$\pm$0.052 & 0.034$\pm$0.019  \\
N 22 &  2 & 11.905884 & $-$73.299297 & 1.950$\pm$0.399 &  4.367$\pm$1.034 &  0.000$\pm$0.269 & 0.040$\pm$0.048 & 0.003$\pm$0.022  \\
N 22 &  3 & 11.909261 & $-$73.308280 & 2.895$\pm$0.455 &  3.297$\pm$0.939 &  0.000$\pm$0.379 & 0.031$\pm$0.075 & 0.031$\pm$0.028  \\
N 22 &  4 & 11.923294 & $-$73.294323 & 1.998$\pm$0.295 &  7.586$\pm$1.009 &  0.000$\pm$0.331 & 0.222$\pm$0.057 & 0.046$\pm$0.020  \\
N 22 &  5 & 11.923392 & $-$73.284345 & 6.922$\pm$0.409 & 14.024$\pm$1.014 &  1.669$\pm$0.206 & 0.352$\pm$0.057 & 0.054$\pm$0.023  \\
N 22 &  6 & 11.923489 & $-$73.274367 & 5.214$\pm$0.474 & 11.883$\pm$0.991 &  1.311$\pm$0.409 & 0.301$\pm$0.054 & 0.069$\pm$0.021  \\
N 22 &  7 & 11.926679 & $-$73.303306 & 2.149$\pm$0.448 &  5.763$\pm$1.168 &  0.000$\pm$0.346 & 0.111$\pm$0.065 & 0.025$\pm$0.018  \\
N 22 &  8 & 11.927250 & $-$73.243438 & 5.895$\pm$0.370 & 10.887$\pm$1.138 &  0.000$\pm$0.619 & 0.147$\pm$0.038 & 0.005$\pm$0.020  \\
N 22 &  9 & 11.930068 & $-$73.312289 & 4.740$\pm$0.560 &  5.598$\pm$0.748 &  0.255$\pm$0.267 & 0.081$\pm$0.047 & 0.035$\pm$0.021  \\
N 22 & 10 & 11.930627 & $-$73.252421 & 3.770$\pm$0.355 &  8.918$\pm$0.942 &  0.000$\pm$0.564 & 0.095$\pm$0.051 & 0.023$\pm$0.016  \\
N 22 & 11 & 11.940693 & $-$73.289347 & 4.091$\pm$0.545 & 12.358$\pm$1.020 &  0.650$\pm$0.237 & 0.956$\pm$0.078 & 0.463$\pm$0.028  \\
N 22 & 12 & 11.940781 & $-$73.279369 & 4.818$\pm$0.457 & 13.174$\pm$0.957 &  0.978$\pm$0.311 & 0.410$\pm$0.083 & 0.081$\pm$0.019  \\
N 22 & 13 & 11.940868 & $-$73.269391 & 3.635$\pm$0.536 &  9.436$\pm$1.282 &  0.652$\pm$0.395 & 0.212$\pm$0.067 & 0.056$\pm$0.024  \\
N 22 & 14 & 11.940955 & $-$73.259413 & 4.311$\pm$0.433 &  9.660$\pm$0.727 &  1.009$\pm$0.221 & 0.184$\pm$0.054 & 0.033$\pm$0.028  \\
N 22 & 15 & 11.944087 & $-$73.298330 & 3.994$\pm$0.543 &  6.894$\pm$0.881 &  0.000$\pm$0.212 & 0.571$\pm$0.062 & 0.120$\pm$0.021  \\
N 22 & 16 & 11.944598 & $-$73.238462 & 3.279$\pm$0.394 &  5.651$\pm$0.863 &  0.000$\pm$0.504 & 0.122$\pm$0.054 & 0.052$\pm$0.027  \\
\enddata
\label{tab:allmeas}
\tablecomments{All PAH and line measurements are integrated intensities
in units of $10^{-8}$ W m$^{-2}$ sr$^{-1}$. The full table is published
in the electronic edition of the journal and includes all of the
measurements and uncertainties for the full set of PAH bands and the
neon lines as well as the values after the ``dark settle'' correction.}
\end{deluxetable*}

\section{Results}\label{sec:results}

\subsection{Band Ratios in the SMC}

In the following section we present our results in a series of plots
showing the band ratios determined in each of the extraction regions
shown on Figures~\ref{fig:coverage1},~\ref{fig:coverage2}
and~\ref{fig:coverage3}.  For comparison we also plot band ratios from
two literature samples: 1) the central regions of galaxies in the SINGS
sample without evidence for AGN from \citet{smith07a} and 2) band ratios
measured in a sample of blue compact dwarfs (BCDs) and starbursts from
\citet{engelbracht08}.  These samples are chosen to represent a range of
metallicities and star-formation rates and to have been measured with
the same PAHFIT technique to ensure consistent results.  The
spectroscopic sample of SINGS galaxies without AGN represent galaxies
that are generally more metal-rich than the SMC.  The starburst sample
from \citet{engelbracht08}, on the other hand, span a range of
metallicities, from Milky Way metallicity down to lower than that of the
SMC.  We note that our measurements are on small scales inside
individual star-forming regions, while the \citet{smith07a} and
\citet{engelbracht08} measurements encompass large regions of a galaxy.
Thus, we use the comparison with caution.  However, the results of Paper
I suggest that most of the PAHs in the SMC reside in the vicinity of
star-forming regions where there is molecular gas, and the majority of
the PAH emission arises in dense PDRs, so in terms of the total PAH
emission from the SMC, star-forming regions like those we have mapped
are likely to dominate because of their high luminosities and the low
PAH fraction elsewhere.

On the band ratio plots (Figures~\ref{fig:r77to62_vs_r77to113}
through~\ref{fig:r170to77_vs_r77to113}) we show a separate panel for
each SMC region.  To be conservative given the uncertainties related to
the ``dark settle'' artifact, we use 5$\sigma$ upper limits on the band
strength.  If a given region is a 5$\sigma$ limit in either the
uncorrected or artifact-corrected spectrum, we consider it a
non-detection and use the least stringent limit on the corrected or
uncorrected band ratios on the plot.  Our choice of 5$\sigma$ upper
limits may introduce a bias into our results towards the brighter PAH
emission regions.  Because of the variability in position and time of
the artifact, we cannot make a simple cut to judge the quality of a
spectrum, so we have inspected the spectra by hand and assigned as
``best'' quality (i.e. not strongly affected by the artifact) or
``normal'' quality.  The ``best'' quality points are shown in order to
highlight our highest confidence band ratios least affected by
systematic shifts due to the artifact.  In these figures we show only
the uncorrected ratios (aside from the case of limits as described
above).  To illustrate the systematic effects of the artifact on the
spectra we show an arrow on each panel that connects the weighted
mean\footnote{Throughout this paper, the weighted mean is defined as the
mean weighted by the uncertainties as 1/$\sigma^2$.} of the uncorrected
distribution to the weighted mean of the artifact-corrected
distribution.  In determining the weighted mean we have used all spectra
where all of the relevant PAH features are detected at $>5\sigma$ in a
given region.  For this reason, the centroid of the distribution changes
slightly from figure to figure due to the different combinations of
bands shown. In panel h) of each ratio plot, we show the SINGS and
starburst samples and the weighted means of the SMC regions.  All panels
show dashed lines indicating the weighted mean of the ratio for the
SINGS galaxies without AGN.

In Table~\ref{tab:meanbr} we list the weighted mean band ratio for each
region along with the uncertainty on the mean.  Note that in this table
the band ratios may differ slightly from the values shown on the band
ratio figures.  This is because when calculating the values for the
figures we have done the weighted mean for only those spectra where all
bands shown in the relevant ratios have been detected (e.g. 6.2, 8.6,
7.7 and 11.3 for a plot of 8.6/6.2 versus 7.7/11.3). For the Table, we
include in our weighted means all ratios where the two relevant bands
were detected, so these may differ slightly from those shown on the
figures. 

\begin{deluxetable*}{lccccccccc}
\tablewidth{0pt}
\tabletypesize{\scriptsize}
\tablecolumns{10}
\tablecaption{Mean Band Ratios}
\tablehead{\multicolumn{1}{l}{Ratio} & \multicolumn{1}{c}{N 22} & \multicolumn{1}{c}{N 66} & 
\multicolumn{1}{c}{N 76} & \multicolumn{1}{c}{N 83} & \multicolumn{1}{c}{SW Bar 1} & \multicolumn{1}{c}{SW Bar 3} &
\multicolumn{1}{c}{SMC B1} & \multicolumn{1}{c}{SINGS} & \multicolumn{1}{c}{SB}}
\startdata
      6.2/11.3 & 1.165$\pm$0.032 & 0.807$\pm$0.037 & 1.182$\pm$0.077 & 1.153$\pm$0.064 & 0.882$\pm$0.015 & 1.240$\pm$0.028 & 0.891$\pm$0.020 & 1.197$\pm$0.276 & 1.169$\pm$0.945 \\
       7.7/6.2 & 2.122$\pm$0.071 & 2.599$\pm$0.163 & 2.454$\pm$0.464 & 2.525$\pm$0.146 & 2.525$\pm$0.061 & 2.428$\pm$0.089 & 2.084$\pm$0.075 & 3.590$\pm$0.638 & 4.160$\pm$1.542 \\
      7.7/11.3 & 2.487$\pm$0.072 & 2.253$\pm$0.121 & 2.730$\pm$0.151 & 2.968$\pm$0.118 & 2.256$\pm$0.047 & 3.155$\pm$0.095 & 1.897$\pm$0.045 & 4.176$\pm$0.927 & 4.684$\pm$3.979 \\
       8.6/6.2 & 0.272$\pm$0.022 & 0.381$\pm$0.022 & 0.315$\pm$0.109 & 0.427$\pm$0.032 & 0.426$\pm$0.013 & 0.379$\pm$0.020 & 0.383$\pm$0.013 & 0.630$\pm$0.143 & 0.782$\pm$0.209 \\
      8.6/11.3 & 0.272$\pm$0.034 & 0.329$\pm$0.028 & 0.369$\pm$0.064 & 0.517$\pm$0.034 & 0.385$\pm$0.011 & 0.465$\pm$0.022 & 0.341$\pm$0.017 & 0.723$\pm$0.095 & 0.801$\pm$0.410 \\
      17.0/7.7 & 0.099$\pm$0.004 & 0.108$\pm$0.009 & 0.071$\pm$0.009 & 0.063$\pm$0.006 & 0.127$\pm$0.003 & 0.075$\pm$0.004 & 0.123$\pm$0.014 & 0.119$\pm$0.039 & 1.092$\pm$4.885 \\
     17.0/11.3 & 0.247$\pm$0.007 & 0.277$\pm$0.017 & 0.189$\pm$0.012 & 0.215$\pm$0.017 & 0.281$\pm$0.008 & 0.238$\pm$0.010 & 0.247$\pm$0.025 & 0.475$\pm$0.115 & 1.974$\pm$6.285 \\
NeIII/NeII & 0.541$\pm$0.107 & 5.679$\pm$0.699 & 1.905$\pm$0.260 & 0.840$\pm$0.132 & 0.394$\pm$0.028 & 1.043$\pm$0.233 &  \nodata & 5.148$\pm$4.222 & 3.148$\pm$4.800 \\
      6.2/$\Sigma_{PAH}$ & 0.183$\pm$0.004 & 0.133$\pm$0.008 & 0.186$\pm$0.010 & 0.151$\pm$0.005 & 0.145$\pm$0.002 & 0.169$\pm$0.004 & 0.161$\pm$0.003 & 0.133$\pm$0.019 & 0.109$\pm$0.029 \\
      7.7/$\Sigma_{PAH}$ & 0.394$\pm$0.006 & 0.407$\pm$0.008 & 0.398$\pm$0.024 & 0.405$\pm$0.008 & 0.376$\pm$0.005 & 0.428$\pm$0.008 & 0.347$\pm$0.006 & 0.466$\pm$0.038 & 0.419$\pm$0.097 \\
      8.6/$\Sigma_{PAH}$ & 0.045$\pm$0.005 & 0.052$\pm$0.003 & 0.053$\pm$0.013 & 0.067$\pm$0.004 & 0.063$\pm$0.002 & 0.061$\pm$0.003 & 0.062$\pm$0.003 & 0.082$\pm$0.014 & 0.081$\pm$0.017 \\
     11.3/$\Sigma_{PAH}$ & 0.149$\pm$0.002 & 0.162$\pm$0.005 & 0.139$\pm$0.003 & 0.120$\pm$0.003 & 0.161$\pm$0.002 & 0.134$\pm$0.002 & 0.181$\pm$0.002 & 0.118$\pm$0.022 & 0.126$\pm$0.071 \\
     12.6/$\Sigma_{PAH}$ & 0.066$\pm$0.002 & 0.065$\pm$0.003 & 0.075$\pm$0.003 & 0.067$\pm$0.002 & 0.056$\pm$0.002 & 0.055$\pm$0.002 & 0.068$\pm$0.002 & 0.064$\pm$0.007 & 0.079$\pm$0.030 \\
     17.0/$\Sigma_{PAH}$ & 0.037$\pm$0.001 & 0.047$\pm$0.003 & 0.028$\pm$0.001 & 0.028$\pm$0.002 & 0.046$\pm$0.001 & 0.033$\pm$0.001 & 0.044$\pm$0.005 & 0.055$\pm$0.015 & 0.132$\pm$0.154
\enddata
\label{tab:meanbr}
\tablecomments{The \citet{engelbracht08} sample means are listed under ``SB'' in the table. All ratios are ratios of integrated intensities in units of W m$^{-2}$ sr$^{-1}$.}
\end{deluxetable*}

\subsubsection{The 6.2, 7.7, 8.6 and 11.3\micron\ Bands}

\begin{figure*}
\centering
\epsscale{1.1}
\plotone{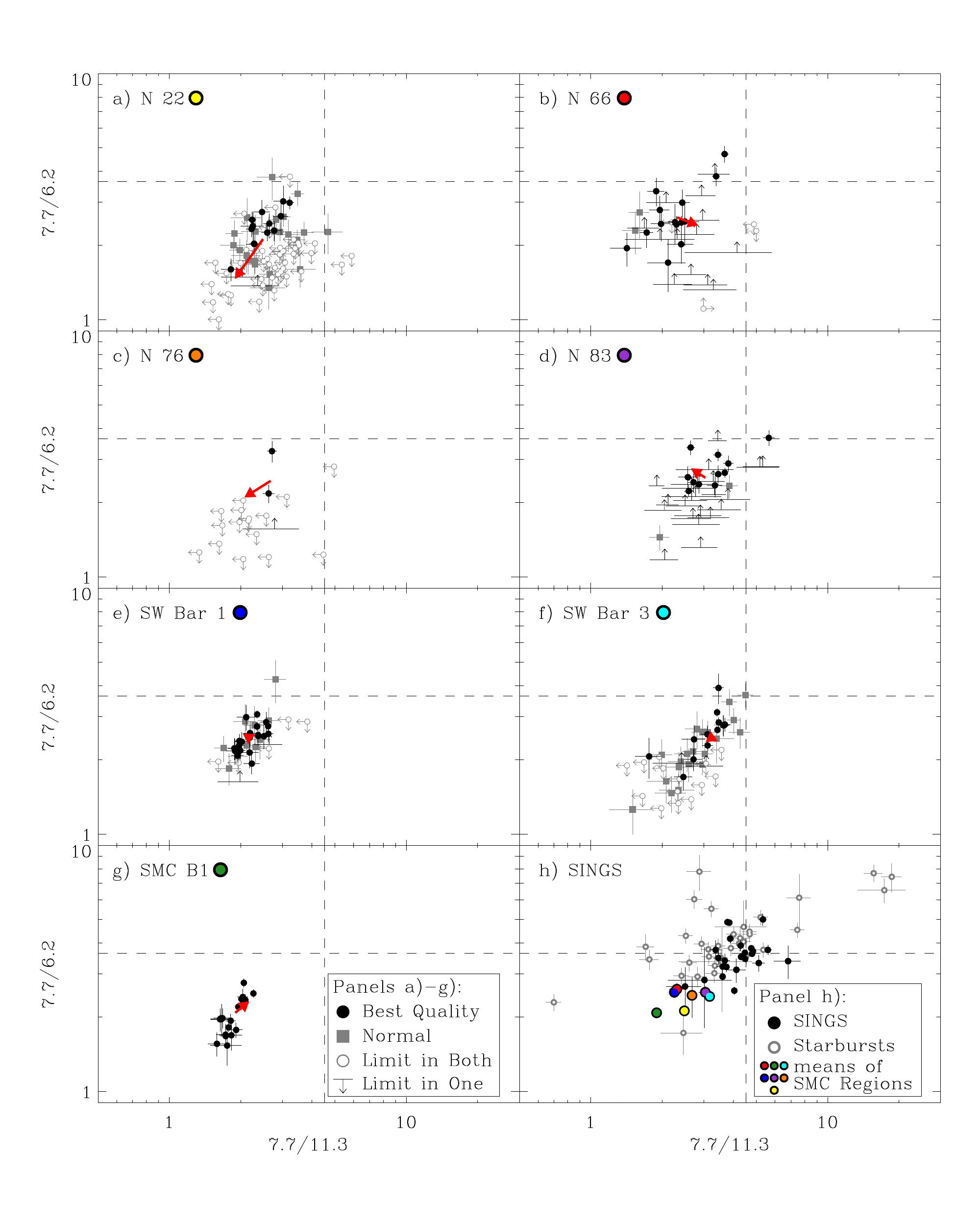}
\caption{The 7.7/6.2 ratio versus the 7.7/11.3 ratio.  The symbols for
each region are indentified in the bottom-right corner of panel g.  A
value is considered a limit in this plot if either the uncorrected or
artifact-corrected spectrum does not have a detection of all bands at
$>5\sigma$.  In that case, we use the least stringent of the corrected
or uncorrected ratios to be conservative.  The ``best'' and ``normal''
quality points are assigned via visual inspection of the spectra to give
some idea of which points are strongly affected by the artifact.  The
red arrow connects the weighted mean of the uncorrected band ratios to
the same value for the corrected ratios.  These weighted means are shown
also in panel h) with their colors identified by a representative point
in each panel a) through g).  The dashed line shows the weighted mean
values for the SINGS galaxies.  Panel h) shows the ratios for the SINGS
sample without AGN and the starbursts from \citet{engelbracht08}.}
\label{fig:r77to62_vs_r77to113}
\end{figure*}

\begin{figure*}
\centering
\epsscale{1.1}
\plotone{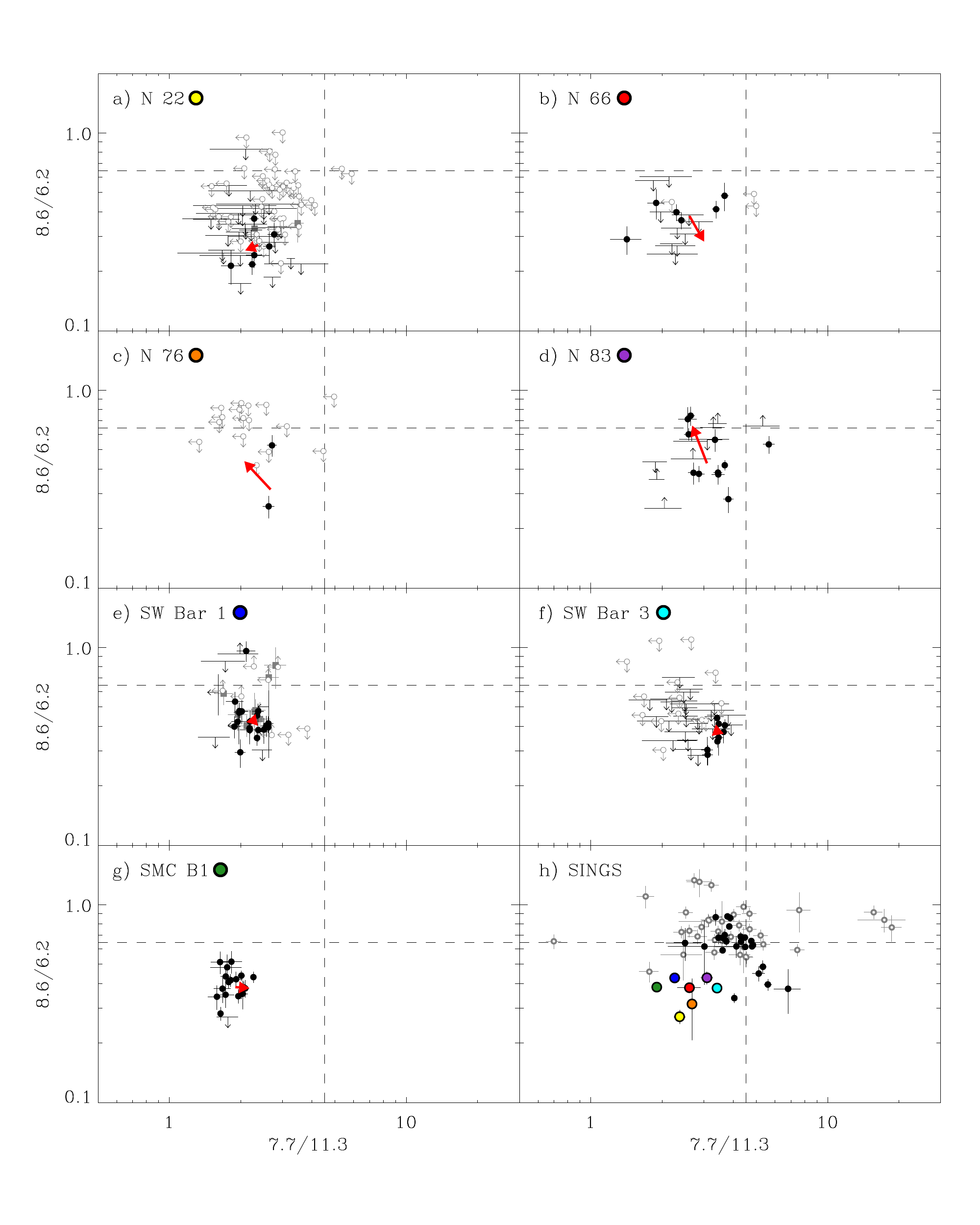}
\caption{The 8.6/6.2 ratio versus the 7.7/11.3 ratio.  The symbols and
annotations in this plot are identical to those in
Figure~\ref{fig:r77to62_vs_r77to113}.}
\label{fig:r86to62_vs_r77to113}
\end{figure*}

\begin{figure*}
\centering
\epsscale{1.1}
\plotone{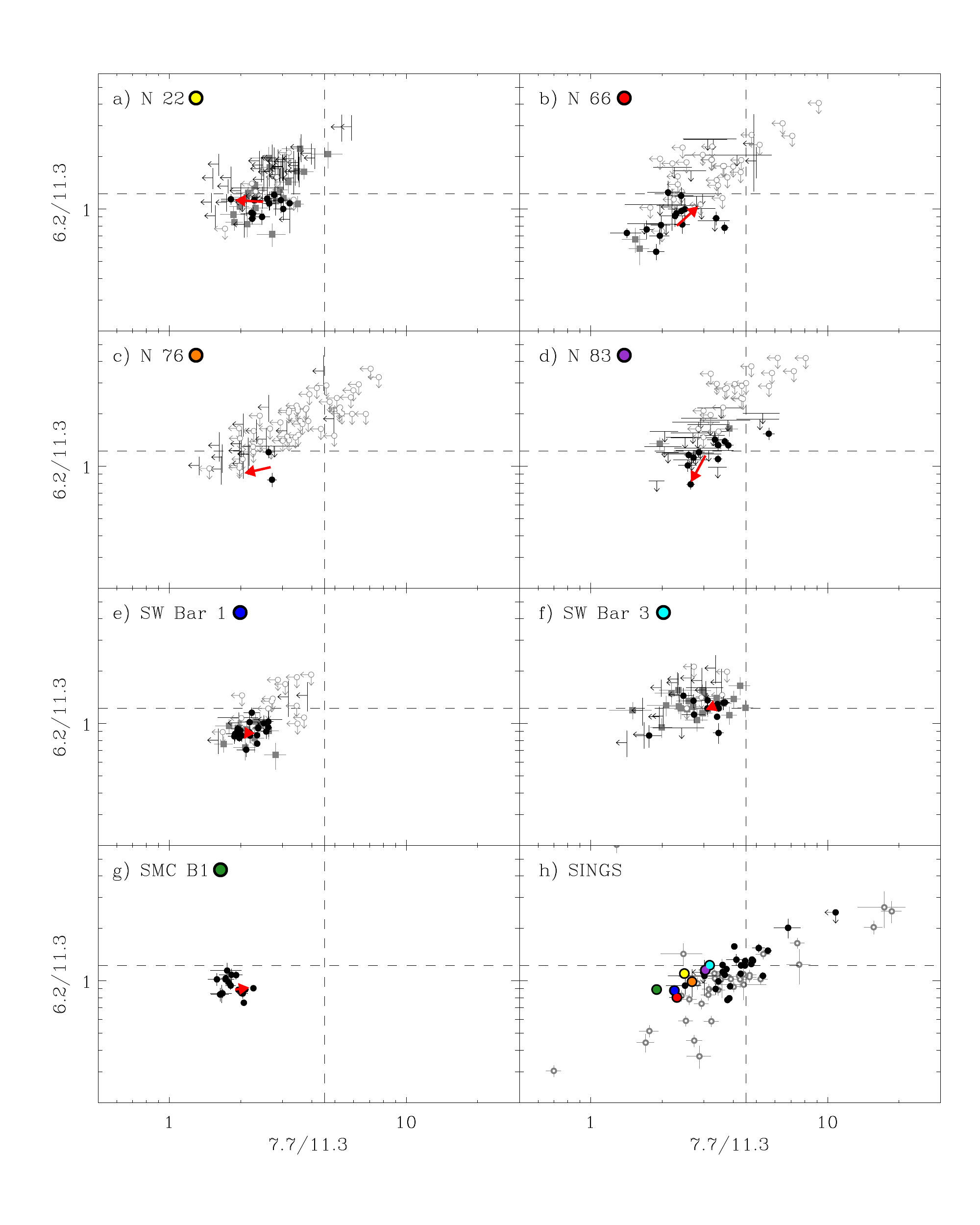}
\caption{The 6.2/11.3 ratio versus the 7.7/11.3 ratio.  The symbols and
annotations in this plot are identical to those in
Figure~\ref{fig:r77to62_vs_r77to113}.}
\label{fig:r62to113_vs_r77to113}
\end{figure*}

\begin{figure*}
\centering
\epsscale{1.1}
\plotone{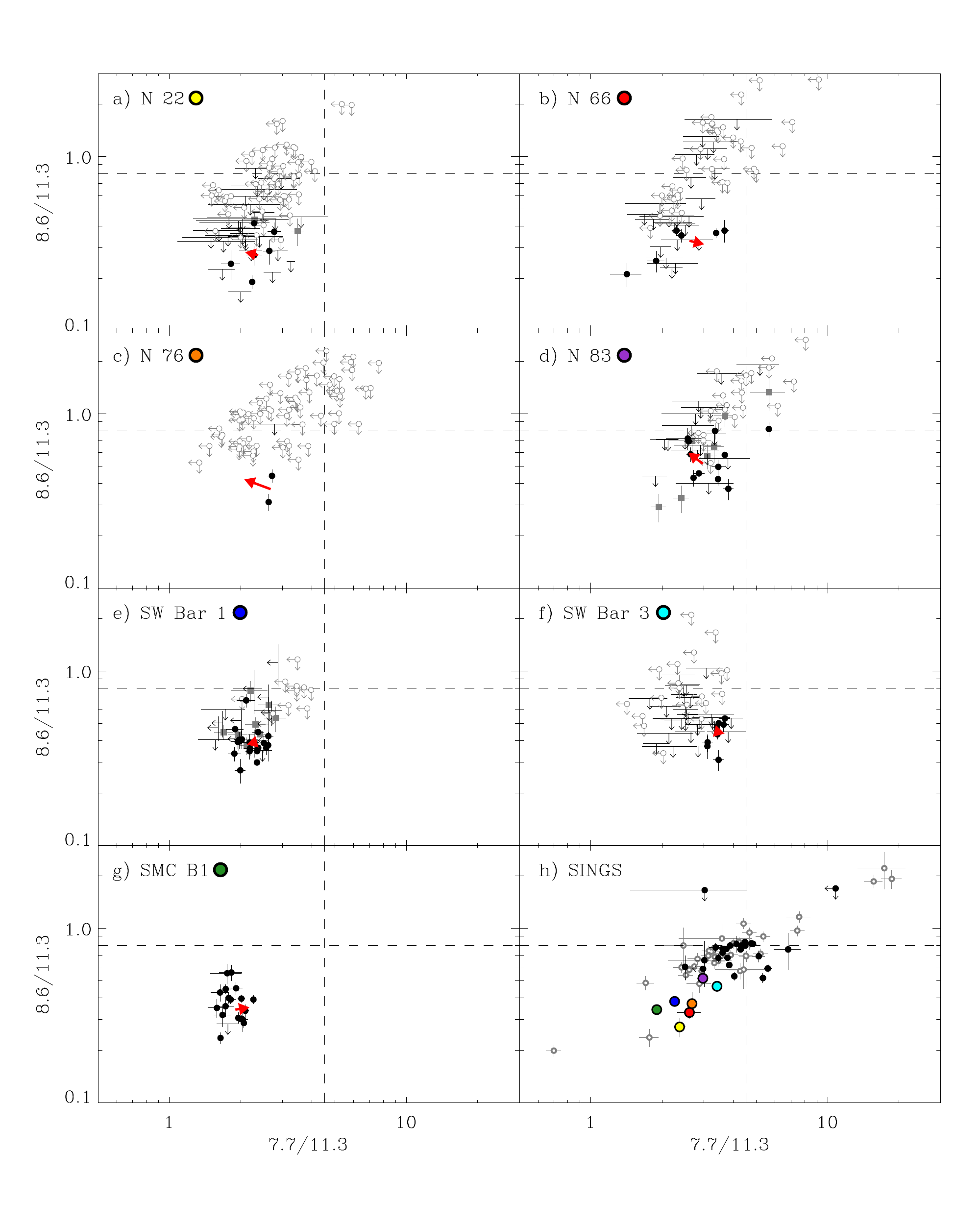}
\caption{The 8.6/11.3 ratio versus the 7.7/11.3 ratio.  The symbols and
annotations in this plot are identical to those in
Figure~\ref{fig:r77to62_vs_r77to113}.}
\label{fig:r86to113_vs_r77to113}
\end{figure*}

Figures~\ref{fig:r77to62_vs_r77to113}, ~\ref{fig:r86to62_vs_r77to113},
~\ref{fig:r62to113_vs_r77to113} and~\ref{fig:r86to113_vs_r77to113} show
the flux ratios of the major PAH bands at 6.2, 7.7, 8.6 and 11.3
\micron.   In these Figures we plot the other ratios versus the
7.7/11.3 ratio, since it generally has the highest signal-to-noise.  We
can see from Figures~\ref{fig:r77to62_vs_r77to113}
through~\ref{fig:r86to113_vs_r77to113} that the 7.7/11.3 is consistently
low in the SMC compared to the SINGS and starburst samples with typical
values of $\sim 2-3$ compared to the SINGS average of $\sim 4-5$.  In
fact, almost all of the individual extracted spectra show 7.7/11.3
ratios below the SINGS average, reinforcing this conclusion.  The ratios
of the 7.7 and 8.6 features to the 6.2 feature are also distinctively
low in the SMC regions compared to the SINGS and starburst samples:
Figures~\ref{fig:r77to62_vs_r77to113} shows that the 7.7/6.2 ratio is
low in essentially all SMC regions and
Figure~\ref{fig:r86to62_vs_r77to113} shows similar results for the
8.6/6.2 ratio.  

In Figure~\ref{fig:r62to113_vs_r77to113} we show the 6.2/11.3 ratio
versus 7.7/11.3. Interestingly, the 6.2/11.3 ratio in the SMC is much
closer to, although still slightly below, the SINGS average.  Comparing
this result with the previous ratios involving the 6.2 band, it appears
that the low ratios of 7.7/11.3 and 7.7/6.2 are mostly due to weakness
of the 7.7 feature and not unusual strengths in the 6.2 or 11.3
features.  We will return to this subject later in the discussion.
Finally, Figure~\ref{fig:r86to113_vs_r77to113} shows consistently low
ratios of 8.6/11.3, reinforcing the weakness of the 8.6 feature in the
SMC regions.

To summarize our findings on the 6.2, 7.7, 8.6 and 11.3 \micron\
features: the SMC regions are consistently lower than the SINGS average
in the 7.7/11.3, 8.6/11.3, 7.7/6.2 and 8.6/6.2 ratios.  The 6.2/11.3
ratio is only slightly lower than the SINGS average, suggesting that the
band ratios involving the 7.7 and 8.6 \micron\ features are low because
of the weakness of those bands (conversely, however, it could be argued
that the 6.2 and 11.3 \micron\ features are abnormally strong).  We note
that the correction of the ``dark settle'' artifact does not strongly
influence these conclusions and in some cases makes the ratios even more
extreme compared to the SINGS average.  We note that if extinction were
a major concern, the 11.3 \micron\ feature would be suppressed, leading
to even more extreme band ratios compared with the SINGS sample.

\subsubsection{The 17.0\micron\ Band}

\begin{figure*}
\centering
\epsscale{1.1}
\plotone{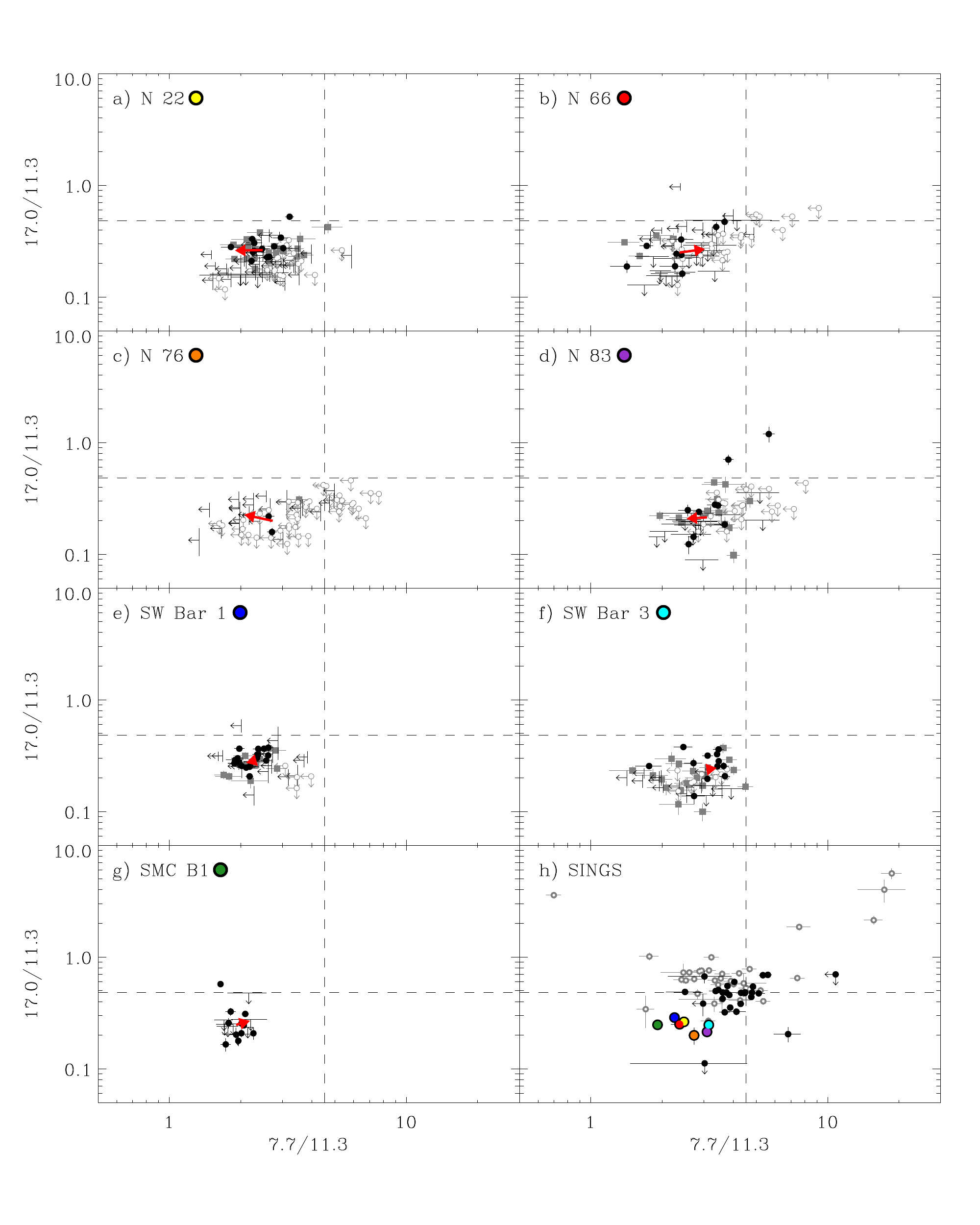}
\caption{The 17.0/11.3 ratio versus the 7.7/11.3 ratio.  The symbols and
annotations in this plot are identical to those in
Figure~\ref{fig:r77to62_vs_r77to113}.}
\label{fig:r170to113_vs_r77to113}
\end{figure*}

\begin{figure*}
\centering
\epsscale{1.1}
\plotone{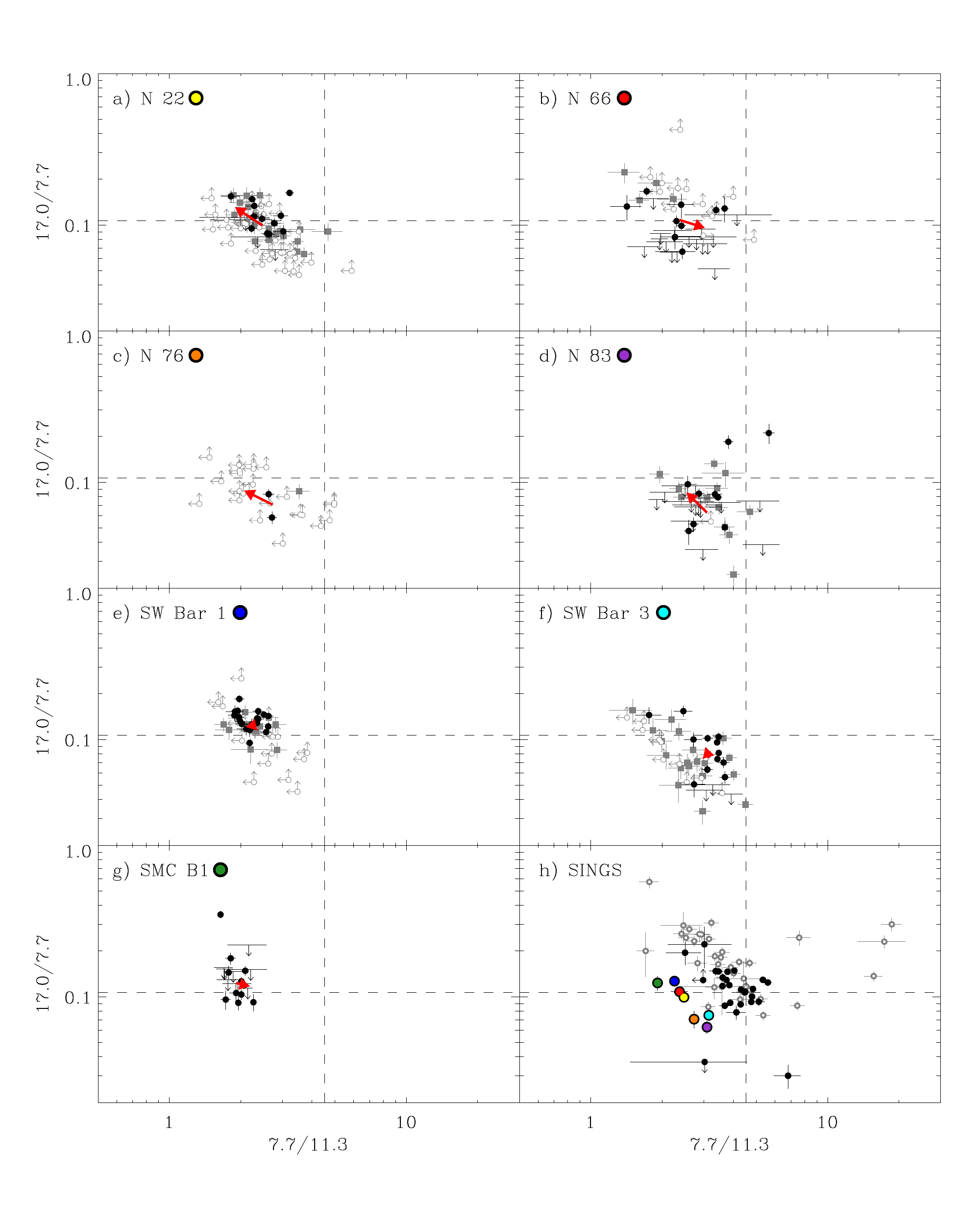}
\caption{The 17.0/7.7 ratio versus the 7.7/11.3 ratio.  The symbols and
annotations in this plot are identical to those in
Figure~\ref{fig:r77to62_vs_r77to113}.}
\label{fig:r170to77_vs_r77to113}
\end{figure*}

In Figures~\ref{fig:r170to113_vs_r77to113}
and~\ref{fig:r170to77_vs_r77to113} we show the ratios of the 17.0
\micron\ band to the 7.7 and 11.3 \micron\ bands.
Figure~\ref{fig:r170to113_vs_r77to113} shows that the 17.0/11.3 ratio is
consistently low in the SMC regions  compared to the SINGS mean.  Aside
from a few outliers, all of the galaxies from SINGS and the starburst
samples which cover the same range of 7.7/11.3 have higher ratios of
17.0/11.3.  There does not appear to be a clear trend in the 17.0/11.3
vs 7.7/11.3 ratio, particularly since the starburst sample has galaxies
with similar 17.0/11.3 ratios but 7.7/11.3 ratios spanning more than an
order of magnitude.  Because the charge state of the carrier of the 17.0
\micron\ complex is not well-constrained, we also show the ratio of
17.0/7.7 in Figure~\ref{fig:r170to77_vs_r77to113}. The region averages
of the 17.0/7.7 ratio in the SMC are closer and at times above the SINGS
average, spanning a range of $\sim$0.06$-$0.1.  We note that for these
ratios, the artifact correction has a minimal effect on the results.

To summarize, we find that the 17.0/11.3 ratio is low in all SMC regions
while the 17.0/7.7 ratio is approximately normal compared to the SINGS
sample mean.  

\subsubsection{Ratios of Individual Bands to the Total PAH Emission}

\begin{figure*}
\centering
\epsscale{0.9}
\plotone{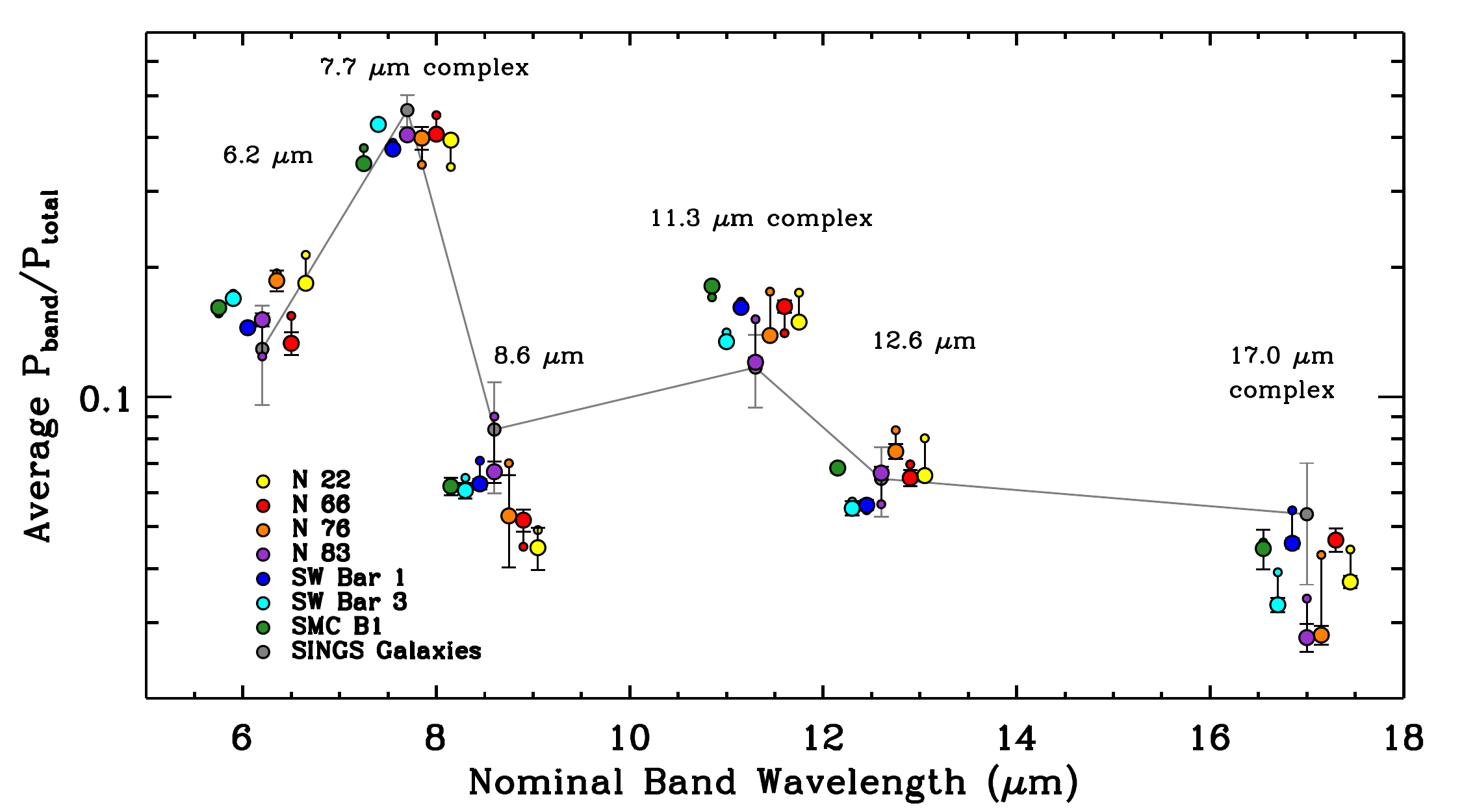}
\caption{The fraction of the total PAH emission carried by each PAH
band.  The colored points represent the weighted mean of the SMC
regions. Small x-axis shifts have been introduced to ensure the points
do not overlap.  These are connected to a smaller point of the same
color to represent the effect of correcting the ``dark settle'' artifact
on the ratio.  The SINGS mean is shown in gray connected with a line to
guide the eye.  The error bar on the SINGS point represents the scatter
of the SINGS galaxies, while the error bars on the SMC points show the
uncertainty in the weighted mean.  Here we omit the 8.3, 12.0, 13.6 and
14.2 \micron\ bands since they make very small contributions to the
total PAH emission.}
\label{fig:ptot}
\end{figure*}

Due to the complexity of comparing among individual bands which may vary
independently, we attempt to simplify the comparison by showing the
ratio of each band to the total PAH emission.  In Figure~\ref{fig:ptot}
we show the fraction of the total PAH emission (defined as the sum of
the 6.2, 7.7, 8.3, 8.6, 11.3, 12.0, 12.6, 13.6 and 17.0 \micron\
features) carried by each of the major PAH features discussed above.  We
show the weighted mean of the ratio for all of the spectra in a given
region where the relevant feature is detected at $>5\sigma$ in both the
original and artifact-corrected version of the spectrum.  The large
points show the weighted mean for the original data and are connected to
a smaller point showing the same weighted mean using the
artifact-corrected spectra.  We overplot the mean from the SINGS sample
(with an error bar showing the standard deviation of the SINGS galaxies)
connected with a gray line to guide the eye.  We can see from this plot
that there are deviations from the SINGS average: 1) the 7.7\ micron\
complex is weaker, 2) the 6.2 and 11.3 \micron\ features are stronger
and 3) the 8.6 and 17.0 \micron\ bands are weaker compared to the SINGS
average.  We note that the conclusion about the weakness of the 17.0
\micron\ complex is enforced by the fact that it is not a major
contributor to the total PAH emission, and therefore should be less
sensitive to changes in the relative proportion of the flux carried by
the 6.2, 7.7, 8.6 and 11.3 \micron\ bands.  This agrees with the
assertion that the weakness of the 7.7 \micron\ feature drives the
relatively normal 17.0/7.7 ratios seen in the SMC.

As described in the Introduction, the C-H bands
between 11$-$14 \micron\ have the potential to provide information on
the structure and hydrogenation of the PAHs.  To compare the relative
strengths of these bands, we make a plot very similar to
Figure~\ref{fig:ptot} but showing the contribution of the 11.3, 12.0,
12.6 and 13.6 \micron\ bands to the total C-H band emission (the sum of
the emission in these features).  For the SMC regions we can see from
this comparison that the 11.3 \micron\ feature carries a similar amount
of the total C-H band emission as in the SINGS galaxies, while the 12.6
\micron\ feature carries slightly less and the 12.0 and 13.6 \micron\
features are slightly stronger. 

\begin{figure*}
\centering
\epsscale{0.9}
\plotone{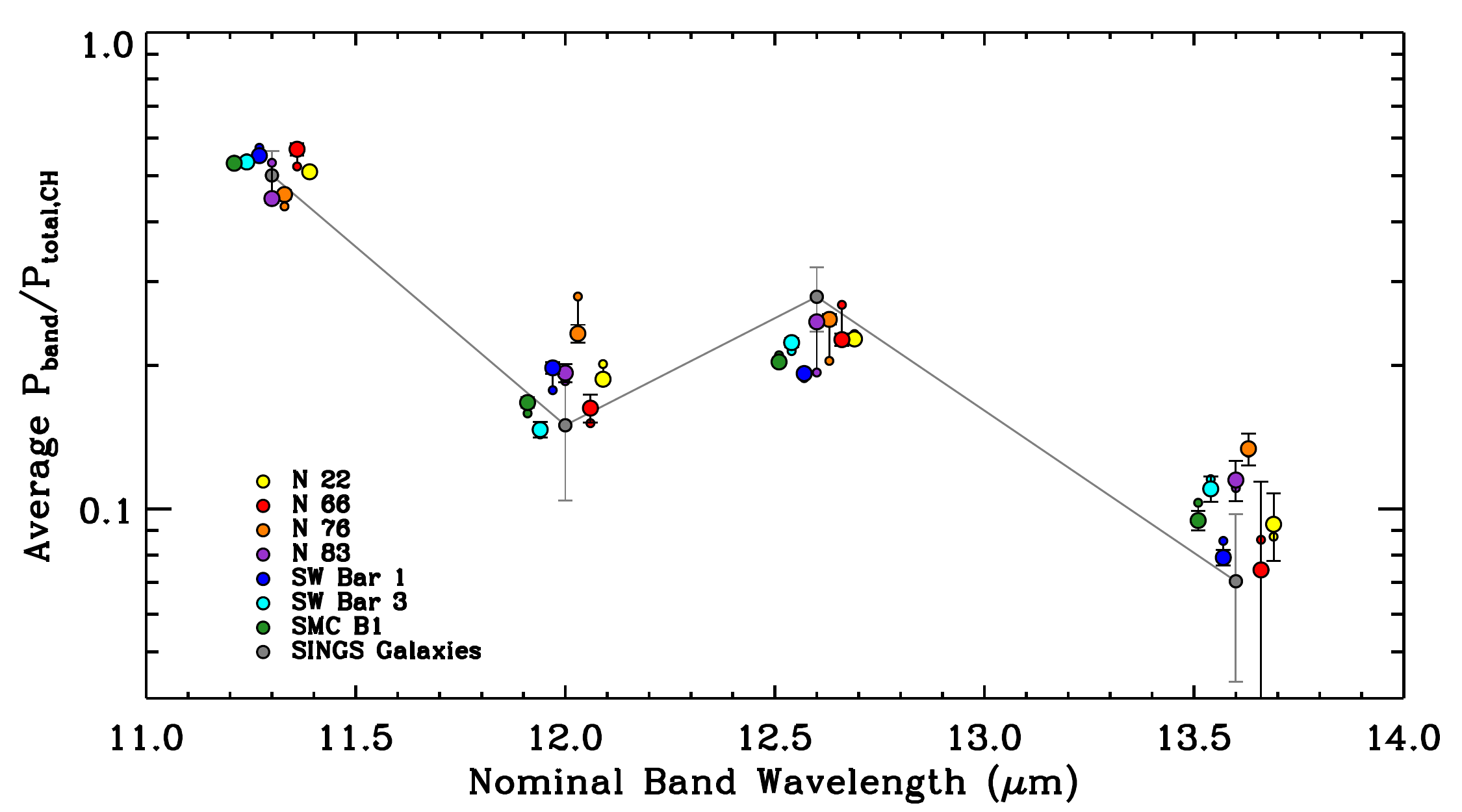}
\caption{The fraction of the total C-H PAH emission in the bands between
11$-$14 \micron\ carried by each individual C-H band.  The colors and
lines are identical to those in Figure~\ref{fig:ptot}.  The 11.3 feature
fraction of the total C-H emission is very similar to that in the SINGS
galaxies, in contrast to the 11.3 feature fraction of the {\em total}
PAH emission, which is higher in the SMC.}
\label{fig:chtot}
\end{figure*}

\subsubsection{Summary of the Band Ratio Results}\label{sec:brsummary}

We have observed clear deviations of the SMC band ratios from the
average values found in the SINGS sample:

\begin{enumerate}

\item For the three brightest PAH bands at 6.2, 7.7 and 11.3 \micron, we
      find that the 7.7 \micron\ feature is weak compared those at 6.2
      and 11.3 \micron.  The 6.2 \micron\ band is slightly weak compared
      to the 11.3 \micron\ band.

\item The 8.6 \micron\ feature appears to be weak relative to all of the
      major bands (6.2, 7.7 and 11.3 \micron) and a weaker contributor
      overall to the total PAH emission.

\item The 17.0 \micron\ feature is weak relative to the 6.2 and 11.3
      features. Compared to the SINGS average, however, the 17.0/7.7
      ratio is only slightly low. The 17.0 \micron\ feature makes a
      smaller contribution to the total PAH emission in the SMC than in
      the SINGS galaxies.

\end{enumerate}

\subsection{Band Ratios and Radiation Field Hardness}

\begin{figure*}
\centering
\epsscale{1.1}
\plotone{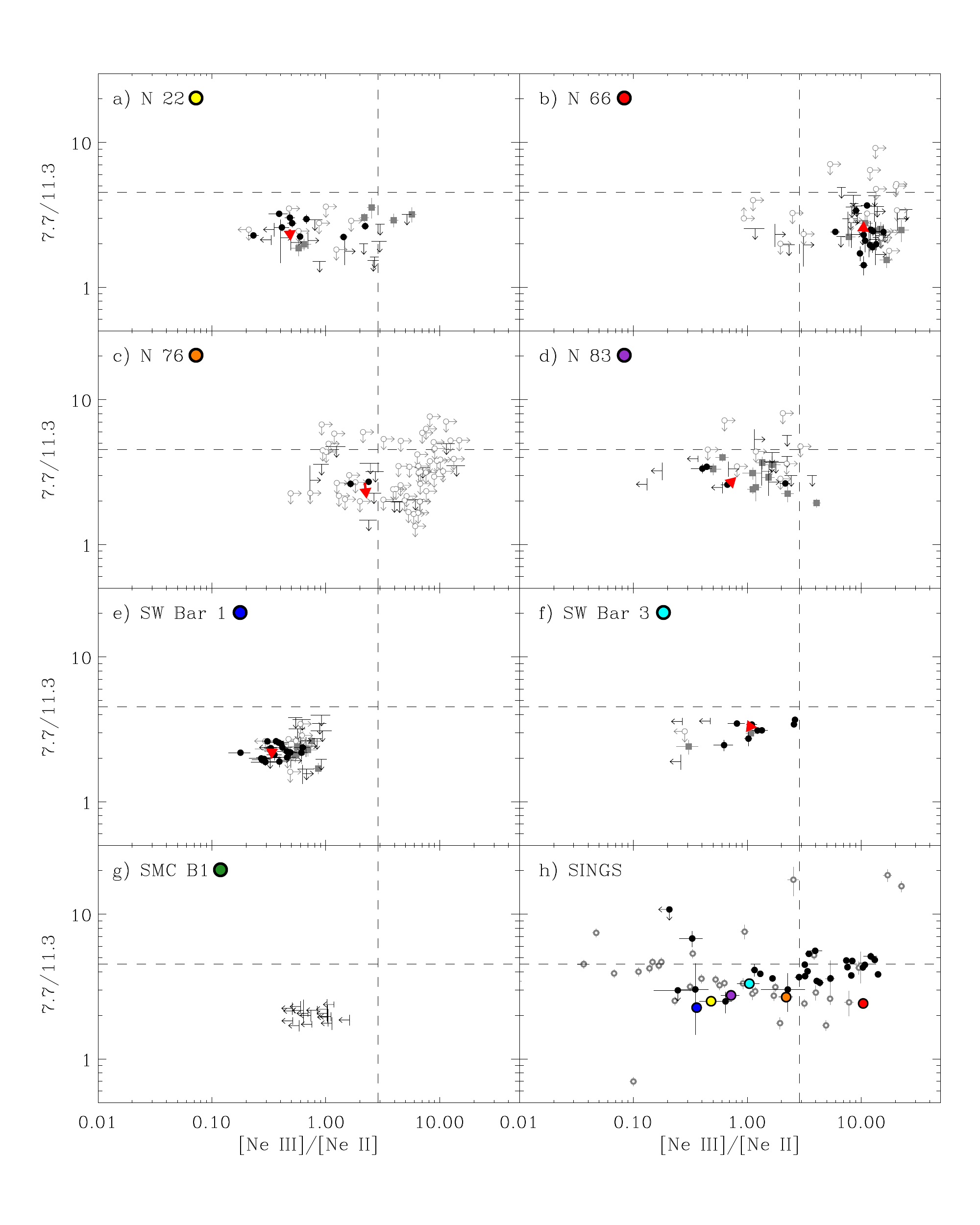}
\caption{The 7.7/11.3 ratio versus radiation field hardness as traced by
the [Ne III]/[Ne II] ratio.  The symbols and annotations in this plot
are identical to those in Figure~\ref{fig:r77to62_vs_r77to113}.  Note
that SMC B1 has no detections of either the [Ne III] or [Ne II] line so
it does not appear in Panel h.}
\label{fig:r77to113_vs_neratio}
\end{figure*}

\begin{figure*}
\centering
\epsscale{1.1}
\plotone{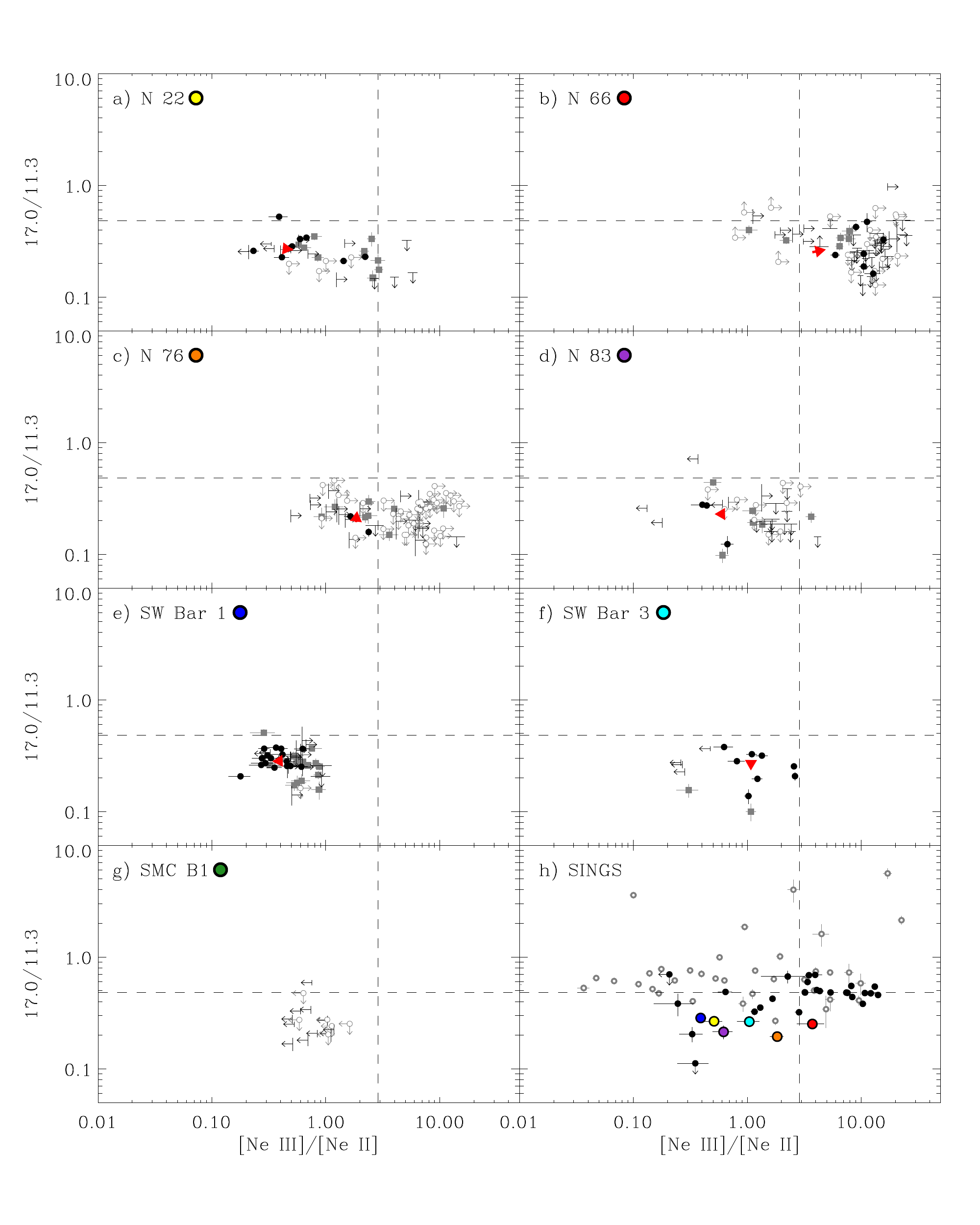}
\caption{The 17.0/11.3 ratio versus radiation field hardness as traced
by the [Ne III]/[Ne II] ratio.  The symbols and annotations in this plot
are identical to those in Figure~\ref{fig:r77to62_vs_r77to113}. Note
that SMC B1 has no detections of either the [Ne III] or [Ne II] line so
it does not apper in Panel h.}
\label{fig:r170to113_vs_neratio}
\end{figure*}

The ratio of the [Ne III] line at 15.6 \micron\ to the [Ne II] line at
12.8 \micron\ traces the hardness of the radiation field in H II regions
\citep{giveon02} and is often used in the context of studying PAH
emission because of the prominence of these two lines in the mid-IR
\citep{madden00,smith07a,gordon08}.  It has been suggested that the
hardness of the radiation field is the driver of the deficit of PAHs
observed at low-metallicity \citep{madden00,gordon08} via enhanced
photodissociation of the molecules.  In order to understand how the
radiation field hardness affects the PAH physical state in the SMC, we
show the 7.7/11.3 and 17.0/11.3 ratios as a function of the [Ne III]/[Ne
II] ratio in Figures~\ref{fig:r77to113_vs_neratio}
and~\ref{fig:r170to113_vs_neratio}.  As will be discussed further in
Section~\ref{sec:discussion} these two ratios predominantly trace
changes in the ionization and size distribution of PAHs, respectively.
We choose these as representative ratios for changes in the PAH physical
state.  We note that none of the other ratios discussed here show trends
with radiation field hardness.

From the ratios shown in these figures, we can see that there is no
clear trend in either the 7.7/11.3 or 17.0/11.3 with radiation field
hardness. The 7.7/11.3 ratio stays essentially constant to within $\sim
50$\% over nearly two orders of magnitude in [Ne III]/[Ne II] ratios.
In fact, we have found no clear trends of any band ratio with the [Ne
III]/[Ne II] ratio.  The same is true for the SINGS galaxies with no
AGN, as seen by \citet{smith07a}.  \citet{brandl06} similarly found no
trend of the 7.7/11.3 ratio over an order-of-magnitude in the neon line
ratio for a sample of starburst galaxies observed with IRS on Spitzer.
In a resolved study of H II regions in M 101, \citet{gordon08} found no
correlation between their ionization index (a combination of the neon
ratio and the ratio of mid-IR silicon lines) and the 7.7/11.3 ratio,
though they do find that the PAH feature equivalent width depends on
radiation field hardness.  These consistent results across a wide range
of physical conditions and metallicity indicate that the radiation field
hardness does not have a strong effect on the PAH band ratios.

\section{The Physical State of SMC PAHs: Interpreting the Band
Ratios}\label{sec:interp}

In Section~\ref{sec:results}, we have enumerated the variations and
average properties of the PAH band ratios we observe in the SMC relative
to galaxies in the SINGS sample.  In the following discussion, we use
information gathered from laboratory and theoretical studies to
translate the ratios into information about the physical state of the
PAHs.  

\subsection{Interpreting the SMC Results}

Our major conclusions about the band ratios in the SMC relative to SINGS
are the following: 1) weak 7.7 \micron\ feature relative to the 6.2 and
11.3 \micron\ features, 2) weak 8.6 \micron\ feature relative to all
major bands and the total PAH emission and 3) weak 17.0 \micron\ feature
relative to the total PAH emission.  

\begin{figure*}
\centering
\epsscale{0.9}
\plotone{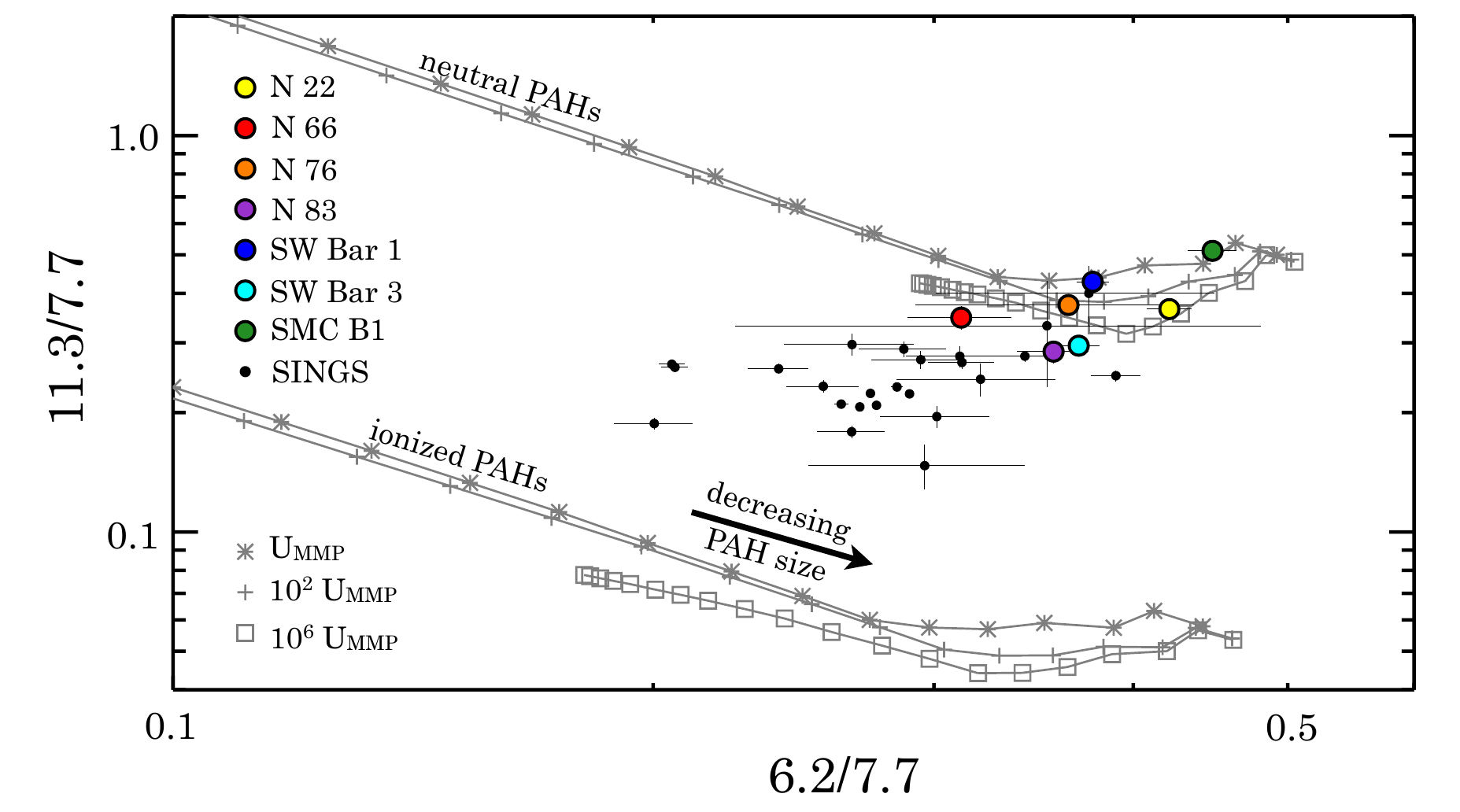}
\caption{Model PAH band ratios from \citet{draine01} with the SINGS
galaxies without AGN overplotted in black and the weighted means of the
SMC regions shown in color.  The two sets of gray lines show the band
ratios for neutral and ionized PAHs and each at three different
radiation field strengths \citep[parametrized in units of
$\mathrm{U}_{\mathrm{MMP}}$, the solar neighborhood radiation field
from][]{mathis83}.  The SMC regions fall in the area of this diagram
compared to the SINGS galaxies that suggests smaller and more neutral
PAHs.}
\label{fig:dlplot}
\end{figure*}

We first address conclusion 1), regarding the ratios of the 6.2, 7.7 and
11.3 \micron\ bands.  As discussed above, the 6$-$9 \micron\ to 11.3
\micron\ band ratios should simultaneously trace changes in the PAH
ionization and size distribution.  To aid in interpreting these ratios
we use a diagram based on the model of \citet{draine01}.  This model
utilizes cross sections for ionized and neutral PAHs as a function of
size based on laboratory and theoretical results.  The PAHs are exposed
to a radiation field with a specified strength and spectral shape and
their emission spectrum is calculated.  Figure~\ref{fig:dlplot} shows
the band ratios from \citet{draine01} for ionized and neutral PAHs at a
range of sizes exposed to radiation fields at 1, 10$^2$ and 10$^6$ times
the \citet{mathis83} solar neighborhood radiation field.  On this plot
we show the band ratios for the SINGS sample with no AGN and the
weighted means for the SMC regions.  We note that the modeled band
ratios shown on this plot are for individual PAHs of a given size, while
PAHs in the ISM should have a wide distribution of sizes.  Therefore, in
comparing the SMC ratios to the models, we use the \citet{draine01}
results as a guide but not as a way to judge specific PAH sizes.

The relative locations of the SINGS galaxies and the SMC regions on this
plot suggest that SMC PAHs may tend to be smaller and more neutral than
those in higher metallicity galaxies.  Within the SMC, N 66 falls
closest to the SINGS mean, suggesting it has relatively larger and more
ionized PAHs.  SMC B1, on the other hand, has band ratios which put it
at the extreme end of the diagram, suggesting it has the smallest and
most neutral PAHs.  The spread in the SMC points may be related to the
progression of the star-formation happening in the region: SMC B1 is a
quiescent molecular region \citep{rubio93,reach00} with abundant
molecular gas, but no detectable H II regions while N 66 has a giant H
II region with multiple generations of star formation in the vicinity
\citep[though it also hosts molecular clouds;][]{rubio00}.  The
remaining regions are intermediate in the star-formation properties but
generally seem to follow a trend where more evolved regions like SW Bar
3 and N 83 are consistent with slightly larger and more ionized PAHs
while regions such as SW Bar 1 and N 22 tend towards smaller and more
neutral PAHs.  

It is worthwhile to note that the \citet{draine01} model attributes all
variation of PAH band strengths to either size or ionization.  It is
very likely that hydrogenation (in excess of the full aromatic
complement of peripheral hydrogens), structure and chemical composition
also play a role in determining the band ratios.  If there were changes
in structure, we would expect to see variations in the relative
strengths of the C-H modes.  More irregular PAHs should show enhanced
emission in the ``duo'', ``trio'' and ``quartet'' modes as discussed in
the Introduction.    Under some conditions, PAHs can become
super-hydrogenated---with two hydrogen atoms attached to a given carbon
atom \citep{lepage03}.  This edge structure with two hydrogens per
carbon is similar to an aliphatic group ($-$CH$_2$) and is expected to
produce similarly aliphatic emission features \citep{bernstein96}.  

Our observations in the SMC do not show strong evidence for changes in
PAH structure (towards more irregular edges) or excess hydrogenation.
In Figure~\ref{fig:chtot} we show the fraction of the 11$-$14 \micron\
PAH emission carried by the 11.3, 12.0, 12.6 and 13.6 \micron\ features.
We note that in the low-resolution spectra, the 12.6 feature is blended
with the [Ne II] line at 12.8 \micron.  For most regions the 11.3 and
12.0 \micron\ bands carry a similar fraction of the C-H flux as seen in
the SINGS sample, however the 13.6 \micron\ feature is enhanced and the
12.6 \micron\ feature is weaker.  Changes in structure that produce more
emission in ``quartet'' modes at 13.6 \micron\ would be expected to also
enhance ``trio'' modes at 12.6 \micron, as both originate in irregular
edges and extensions.  Thus, we can not draw a clear conclusion from
these band ratios.  At the signal-to-noise of our spectra, we do not see
clear evidence for any aliphatic features that might result from excess
hydrogenation.  Clear detection of these features if they exist and
higher signal-to-noise measurements of the 13.6 and 14.2 \micron\ bands
in the future with JWST will enhance our ability to judge whether SMC
PAHs show changes in structure or hydrogen content.

We now move on to conclusions 2) and 3) regarding the weakness of the
8.6 and 17.0 \micron\ features in the SMC.  As discussed in
the Introduction both 8.6 and 17.0 \micron\ features are expected
to arise in large PAHs.  Based on theoretical calculations of PAH
spectra, the 8.6 \micron\ feature is thought to trace mainly charged
PAHs \citep{bauschlicher08}.  There is no clear identification of the
charge state of the 17.0 \micron\ carrier.  The weakness of these two
features in our spectra provide independent evidence for our
interpretation that PAHs in the SMC tend to be smaller and more neutral
than their counterparts in more metal rich systems.  The particular
weakness of the 8.6 \micron\ feature may be due to the combination of
size and charge state both favoring less emission in that band.

\subsection{Trends of PAH Physical State with Radiation Field Hardness}

In Figures~\ref{fig:r77to113_vs_neratio}
and~\ref{fig:r170to113_vs_neratio} we show the variation of the band
ratios as a function of the [Ne III]/[Ne II] ratio.  These lines arise
in ionized gas and their ratio traces the hardness of the ionizing
radiation field.  We find no clear evidence for variation of the band
ratios with radiation field hardness over nearly two orders of magnitude
in the [Ne III]/[Ne II] ratio.  PAHs are observed to be depleted in
ionized gas \citep{povich07,lebouteiller11} and theoretical studies
suggest that the lifetimes of PAHs in ionized gas can be short compared
to the H II region lifetimes \citep{micelotta10b}.  Because of the
destruction of PAHs in ionized gas, we may not expect the physical state
of the PAHs to strongly depend on the neon line ratio since the PAH
emission we observe arises from the PDR around the H II region and is
thus physically separate from the gas traced by the neon lines.

\subsection{Trends of PAH Physical State with Metallicity}

\begin{figure*}
\centering
\epsscale{1.1}
\plottwo{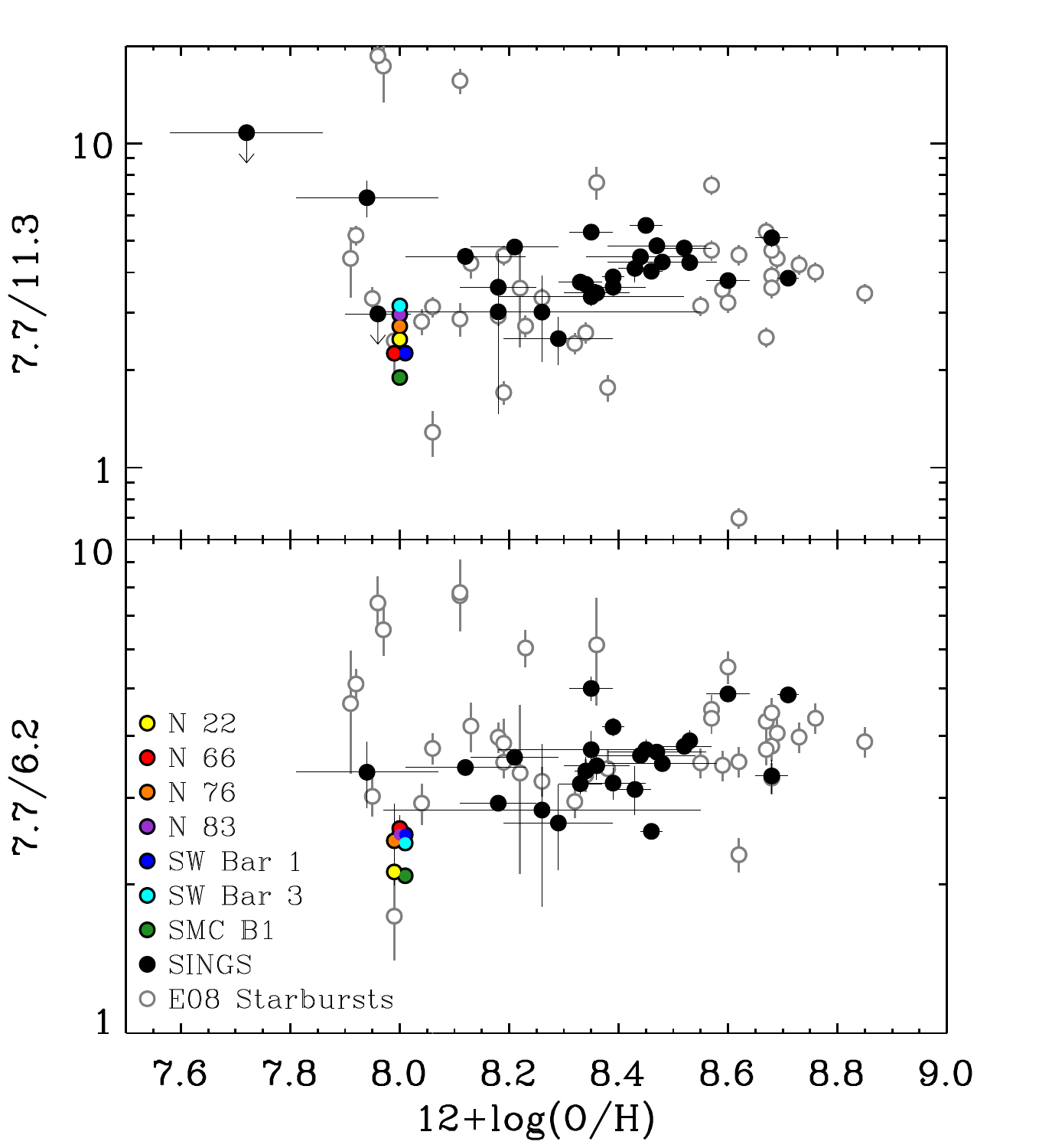}{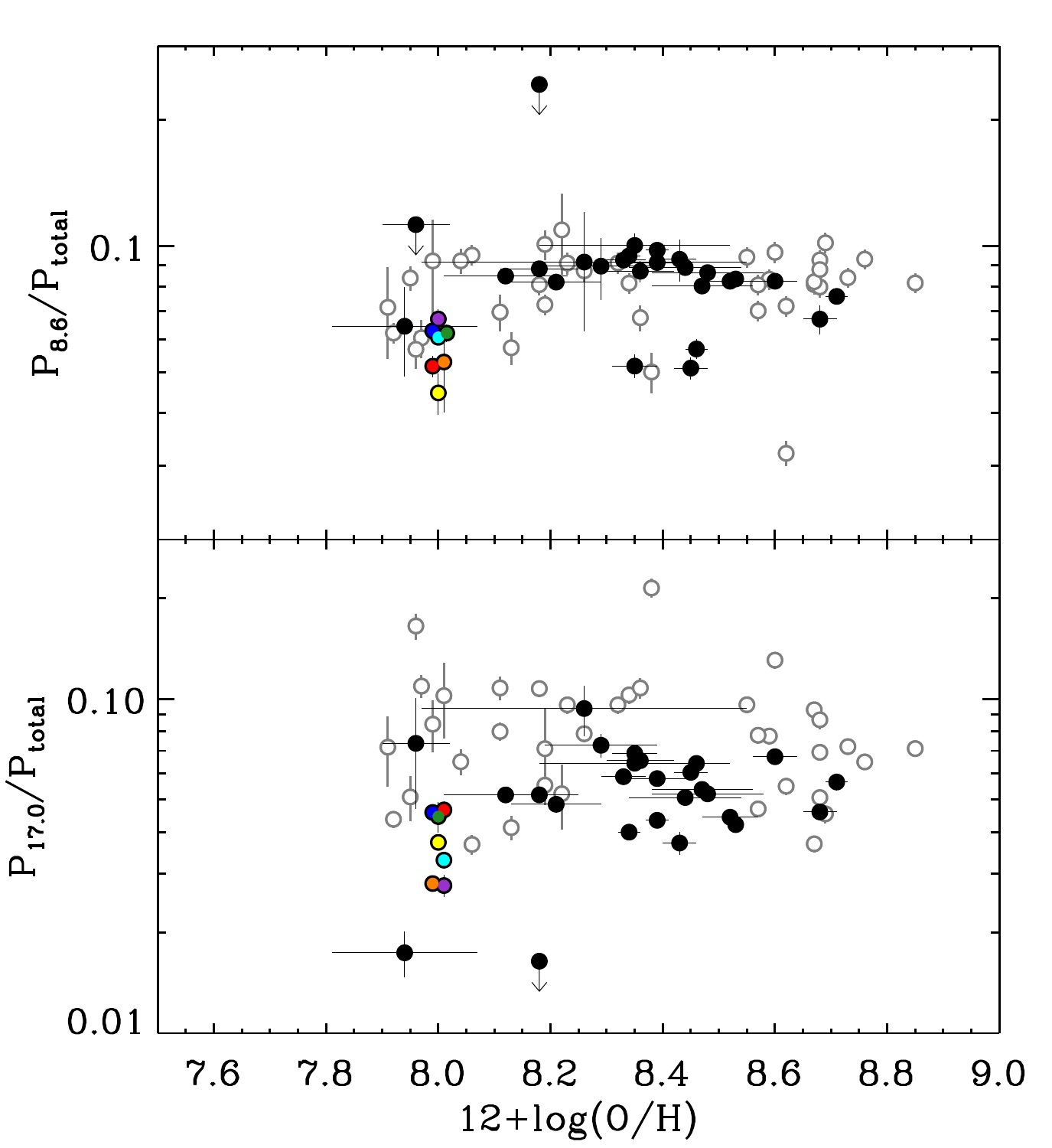}
\caption{The band ratios from the SINGS \citep{smith07a}, starburst
\citep{engelbracht08} and SMC samples versus metallicity.  For the SINGS
galaxies we use the metallicities for the central regions from
\citet{moustakas10} using the calibration of \citet{pilyugin05}.  The
starburst metallicities are T$_{\rm e}$ based and should be consistent with the
\citet{pilyugin05} results.  They are described further in \citet{engelbracht08} and do
not have uncertainties.  We use a metallicity of $12+$log(O/H)$\sim 8.0$
for the SMC \citep{kurt98}.  Small shifts on the x-axis are introduced
for the SMC points so they do not overlap.}
\label{fig:vs_metals}
\end{figure*}

In Figure~\ref{fig:vs_metals} we show a series of plots illustrating the
 variation of the several PAH band ratios with metallicity.  In general,
 the SMC occupies a similar region of the plot as other low-metallicity
 galaxies from the SINGS and starburst samples, though there is
 considerable scatter.  There is a slight trend towards lower 7.7/11.3
 ratios as a function of metallicity, though with many outliers.
 \citet{smith07a} found a slight trend for galaxies to show low
 17.0/11.3 ratios at lower metallicity.  We show their galaxies here,
 instead using the 17.0/$\Sigma_{PAH}$ as a tracer of the 17.0 \micron\
 feature strength that does not depend as heavily on the charge state of
 the PAHs.  The SMC 17.0 \micron\ fraction of the total PAH emission
 falls below the average for SINGS but above the lowest metallicity
 SINGS galaxy with detected PAH emission (NGC 2915).
 
\section{Discussion}\label{sec:discussion}

\subsection{Insights Into the PAH Life-Cycle as a Function of
Metallicity}\label{sec:lc}

By comparing our measured band ratios with laboratory and theoretical
results, we infer that PAHs tend to be smaller on average in the SMC
than in higher metallicity galaxies like those represented by the SINGS
sample.  Note that this is not to say that the SMC is devoid of large
PAHs.  We do see a clear 17.0 \micron\ feature in our spectra, which
requires large PAHs to be present in the ISM at some level.  The band
ratios suggest that the proportion of large PAHs is decreased relative
to small PAHs.  If the SMC represents a typical low-metallicity galaxy,
we argue that the smaller average PAH sizes have two major implications
for the PAH life-cycle at low-metallicity: 1) the change in the size
distribution can not be the product of processing in the ISM but must
instead reflect changes in PAH formation and 2) the deficit of PAHs in
low-metallicity galaxies may be related to the fact that PAHs are formed
with smaller average sizes and are therefore more easily destroyed under
typical ISM conditions.

\subsubsection{Smaller Average PAH Size Due to Formation not Processing}

PAHs can be destroyed in many ways: photodissociation
\citep{allain96a,allain96b}, interactions with energetic particles in
hot and/or shocked gas \citep{micelotta10a,micelotta10b} and interaction
with cosmic rays \citep{micelotta11}.  For all of these processes, our
current knowledge of PAH destruction suggests that small PAHs are
destroyed more readily than large PAHs.  For instance, theoretical
models of PAH photodissociation in the diffuse ISM suggest that PAHs
with $\lesssim50$ C atoms are quickly destroyed under typical
interstellar conditions \citep{allain96a,allain96b,lepage03} while large
PAHs persist longer under the same ISM conditions.  Similar trends hold
true for processing by shocks, cosmic rays and hot gas
\citep{micelotta10a,micelotta10b,micelotta11}. Therefore, since small
PAHs are destroyed more easily than larger PAHs in the ISM, explaining a
deficit of large PAHs cannot be the effect of ISM processing unless the
destruction process turns large PAHs into small PAHs as a side-effect.
Recent studies by \citet{micelotta10a} suggest that this is not the
case.  The dominant destruction mechanism for large PAHs is through
interaction with supernova shocks and they find that ``daughter'' PAHs
formed by the fragmentation of the larger grains are very quickly
destroyed in the shocked gas.  Thus, since small PAHs are destroyed more
easily in all ISM processes and they are likely not replenished by the
fragmentation of larger PAHs, explaining a deficit of large PAHs cannot
be the effect of ISM processing. 

We do, however, have ample evidence that processing in the ISM can
affect the PAH size distribution in the sense of selectively {\em
removing} small PAHs.  In the vicinity of AGN, multiple studies have
shown a decrease in the 6$-$9 \micron\ features relative to the 11.3
\micron\ and longer wavelength features
\citep{smith07a,odowd09,diamond-stanic10,wu10}.  \citet{smith07a} also
found that the 11.3/17.0 ratio decreases in the presence of an AGN,
suggesting a stronger contribution from the 17.0 \micron\ feature---a
tracer of larger PAHs. These trends suggest that the ISM conditions in
the vicinity of the AGN have selectively removed the more fragile,
smaller PAHs.  

An interesting counterpoint to our study of the SMC is that by
\citet{hunt10}, who presented a study of the PAH band ratios in a sample
of blue compact dwarf galaxies (BCDs).  These BCDs have particularly
intense radiation fields as well as low metallicities and very low
dust-to-gas ratios (hence decreased dust shielding).  \citet{hunt10}
found that the band ratios in these galaxies suggested a size
distribution shifted towards larger PAHs relative to the galaxies in the
SINGS sample (with no AGN), particularly in the strength of the 8.6
\micron\ feature and the low ratios of short-to-long wavelength bands.
Because of the low metallicity of these galaxies, we might expect them
to show a size distribution shifted towards smaller PAHs, as we have
seen in the SMC.  It may be the case in these extreme objects that the
radiation field is intense enough to remove small PAHs to such a degree
that once again larger PAHs dominate the size distribution, regardless
of the scarcity of large PAHs.  If this were the case, we would expect
the abundance of PAHs relative to dust (i.e. the PAH fraction) to be
very low in the BCDs.  Indeed, \citet{hunt10} found that the ratio of
PAH emission to the total infrared emission ($\Sigma{\rm PAH}$/TIR),
which should trace the PAH fraction, is depressed by more than an order
of magnitude on average in these galaxies compared to the SINGS sample. 

If processing in the ISM cannot produce the size distribution shifted
towards smaller PAHs that we observe, the only other option is that PAHs
are formed with smaller sizes in the SMC. Forming PAHs with smaller
average sizes in low-metallicity galaxies could be the result of a
change with metallicity in the dominant formation mechanisms (i.e.~AGB
stars, coagulation or chemistry in dense gas---among many other options
that have been proposed) and/or a change in the effectiveness of forming
large PAHs with the same mechanism. Both of these options are plausible,
especially given the simple fact that the raw material that PAHs form
out of is less abundant in low-metallicity galaxies, so constructing
large PAHs may be inherently more difficult.

\subsubsection{Low-Metallicity PAH Deficit as a Consequence of Smaller PAH Sizes}

An important consequence of PAHs forming with a distribution shifted
towards smaller sizes at low-metallicity is that a larger fraction of
PAHs can be destroyed under typical ISM conditions.  In general, models
of PAH destruction in the ISM find that below a certain PAH size
\citep[$\sim 20-50$ C atoms, generally;][]{lepage01,lepage03} PAHs are
very quickly destroyed.  The timescale for destruction of small PAHs is
shorter for essentially all destruction mechanisms, as discussed above.
With a size distribution shifted to smaller sizes, the quickly destroyed
PAHs will represent a larger fraction of the total amount of PAHs.  Such
a situation may naturally lead to a PAH deficit in low-metallicity
galaxies without the need to rely on enhancements in radiation field
hardness \citep{madden00,gordon08} or delays between PAH formation in
AGB stars versus dust formation in supernovae \citep{galliano08b}. We
note, however, various mechanisms can be operating simultaneously---i.e.
PAHs could be both easier to destroy because of their smaller sizes
while also being exposed to harder, more intense radiation fields in
low-metallicity galaxies.  The interaction of these two effects could
contribute to the steepness of the observed drop-off in the PAH fraction
below a metallicity of 12$+$log(O/H)$\sim8.1$
\citep{engelbracht05,draine07b}.

\subsubsection{Relationship of PAH Emission and Radiation Field Hardness}

It has been suggested that the deficiency of PAHs in low metallicity
galaxies is a product of their destruction by harder and/or more intense
radiation fields \citep[see for instance,][]{madden00,gordon08}.  We
have shown that in the SMC the band ratios do not depend strongly on
the hardness of the ionizing radiation field, as one might expect if it
were instrumental in destroying PAHs.  In fact, a wide range of studies
have shown no correlation between the radiation field hardness and the
band ratios \citep{brandl06,smith07a,gordon08,odowd09,wu10}, in line
with these two tracers probing distinct physical regions (H II region
vs. PDR).  Some studies do show, however, a correlation of PAH
equivalent width with radiation field hardness
\citep[e.g.][]{madden00,beirao06,gordon08}, such that PAHs have lower
equivalent widths in harder radiation fields.  This may be the product
of the correlation between harder and more intense radiation fields.
Young, massive stars capable of creating harder ionizing radiation will
also have more intense radiation fields.  The intensity of the radiation
field may then affect the amount of emission arising in PDRs around the
HII region.  Meanwhile, the characteristics of the PAH spectrum are
determined by conditions within the PDR itself and may be insensitive to
the hardness of the ionizing radiation \citep[for a similar argument
see][]{brandl06}.  

\subsubsection{Why are SMC PAHs more Neutral?}

In addition to being smaller on average than PAHs in the SINGS galaxies,
the SMC PAHs also appear to be less ionized. One possible reason for
lower levels of ionization may be that PAHs in the SMC are largely
confined to dense regions \citep{sandstrom10}.  The ratio of the
ionization to recombination rates should determine the PAH charge and is
proportional to $(G_0/n_e)\times \sqrt{T_{gas}}$.  If PAHs in the SMC
exist in denser regions compared to PAHs at higher metallicity, a
decreased ionization level may be a natural consequence.  Another reason
why PAHs in the SMC may be more neutral is related to their smaller
average sizes.  Small ionized PAHs may have a shorter lifetime than
small neutral PAHs, since ionization in general causes PAHs to be less
stable to photodissociation. It is possible that PAHs in the SMC tend to
be more neutral because small, ionized PAHs are destroyed even more
quickly in the ISM than their neutral counterparts.  More detailed PDR
modeling including the PAH ionization state may provide insight into why
the PAH charge state is different in the SMC than in higher metallicity
galaxies.

\subsubsection{A Scenario for the PAH Life-Cycle in the SMC}

\citet{sandstrom10} argued that in the SMC the low mass fraction of dust
in PAHs in the diffuse regions compared to the higher fraction in dense
regions suggested that PAHs must be forming in dense gas.  Their
reasoning was that AGB-produced PAHs would primarily be deposited in the
diffuse ISM which would then condense to form dense gas, so enhancements
of the PAH fraction in dense gas must occur in situ.  We propose the
following (speculative) scenario to jointly explain the abundance and
physical state of SMC PAHs inferred from our study and that of
\citet{sandstrom10}: 1) PAHs are forming in regions of dense gas in the
SMC; 2) these PAHs are formed with a smaller average size possibly due
to the lower abundance of carbon available in the ISM; 3) as they emerge
from dense regions into the diffuse ISM, the PAHs interact with UV
photons, shocks and cosmic rays, and a fraction of the PAHs are
preferentially destroyed on the small end of the PAH size distribution.
In this scenario, a large fraction of the PAHs in the SMC are destroyed
in the diffuse ISM, causing a PAH deficit, not necessarily because of
harsher conditions, but because the average PAH starts off smaller and
thus more ``fragile'' in the SMC than the average PAH in a higher
metallicity galaxy.

\section{Summary \& Conclusions}\label{sec:conclusions}

We have presented a study of the PAH physical state diagnosed through
the ratios of the 6$-$17 \micron\ bands in several regions of the SMC.
In comparing our measured band ratios to the SINGS sample, excluding
galaxies with AGN, we find several distinct trends: SMC PAHs show a weak
7.7 \micron\ feature relative to the other dominant PAH bands at 6.2 and
11.3 \micron\ and the features at 8.6 and 17.0 \micron\ are weaker
contributors to the overall PAH emission.  We find no trend of the band
ratios as a function of radiation field hardness as traced by the mid-IR
[Ne III]/[Ne II] ratio.

Interpreting the band ratios in light of laboratory and theoretical
results on the PAH spectrum, we suggest that SMC PAHs are smaller and
more neutral than their counterparts in higher metallicity galaxies.
Several lines of evidence argue for this interpretation: the 7.7/11.3
and 7.7/6.2 ratio suggests that PAHs are smaller and more neutral based
on comparison with the \citet{draine01} model and various theoretical
and laboratory studies; the 8.6 \micron\ band, which is thought to be
dominantly emitted by large, charged PAHs is particularly weak in the
SMC; and the 17.0 \micron\ feature---which is thought to arise in
skeletal vibrations of large PAHs---is also weak relative to the total
PAH emission in the SMC. 

We argue that a size distribution shifted towards smaller average sizes
can not be the product of ISM processing by UV photons, shocks, cosmic
rays or hot gas, since all of those processes more efficiently destroy
small PAHs.  A variety of literature results suggest that ISM processing
creates a size distribution shifted towards larger PAH sizes, the
opposite trend we see here.  If ISM processing can  not create the size
distribution we observe, the likely explanation is a change in how PAHs
form.  This could be a change in the mechanism of PAH formation at low
metallicity or a change in the efficiency of creating large PAHs.

Our observations imply that the PAH deficiency at low-metallicities may
not be due to enhancements in the hardness or intensity of radiation
fields, but rather that PAHs at low-metallicity start off inherently
more ``fragile'' because of their small sizes so a larger proportion of
them can be destroyed under typical ISM conditions.

\acknowledgements

We thank Chad Engelbracht and Karl Gordon for providing the starburst
sample PAHFIT results, Erik Muller for providing the SMC HI map, and
Carl Starkey and Remy Indebetouw for their help with the ``dark settle''
artifact correction. We thank the anonymous referee for helpful comments
on the structure of the paper.  This work is based on observations made
with the Spitzer Space Telescope, which is operated by the Jet
Propulsion Laboratory, California Institute of Technology under a
contract with NASA. This research was supported in part by NASA through
awards issued by JPL/Caltech (NASA-JPL Spitzer grant 1264151 awarded to
Cycle 1 project 3316, and grants 1287693 and 1289519 awarded to Cycle 3
project 30491). A.B. wishes to acknowledge travel support from
FONDECYT(CHILE) grant No1080335. A. B. wishes to acknowledge partial
support from grants NSF AST-0838178 and NSF AST-0955836, as well as a
Cottrell Scholar award from the Research Corporation for Science
Advancement RCSA 19968. A.L. is supported in part by a NSF grant
AST-1109039. M.R.  wishes to acknowledge support from FONDECYT(CHILE)
grant No1080335 and is supported by the Chilean {\em Center for
Astrophysics} FONDAP No.  15010003. This research has made use of NASA's
Astrophysics Data System.

\appendix

\section{Artifacts in the IRS Spectral Mapping and their Mitigation}

\subsection{Description of the Artifact}

\begin{figure*}
\centering
\epsscale{0.9}
\plotone{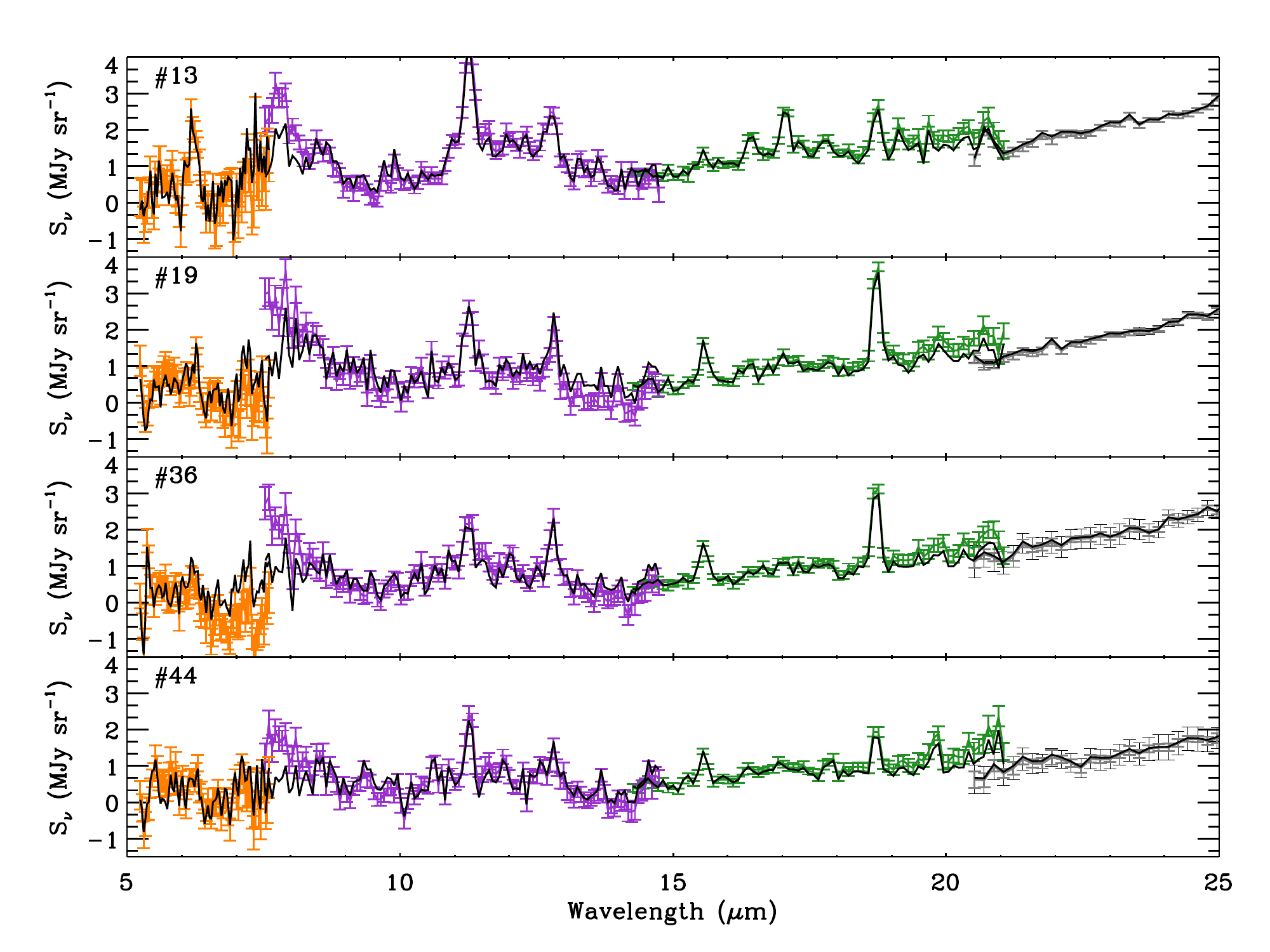}
\caption{Several spectra from the map of SW Bar 1 illustrating various
degrees of mis-match at the SL2/SL1 and LL2/LL1 overlap regions.  The
uncorrected orders are shown in color---SL2 in orange, the SL1 in
purple, the LL2 in green and the LL1 order in gray.  The offsets in the
overlap regions are due to a time- and positionally varying dark level
on the detector.  The black spectrum overlayed shows these same regions
after the correction described in this Appendix.  The correction derived
from the inter-order regions greatly improves the mis-match at the
SL2/SL1 overlap, but it may oversubtract real 7.7 \micron\ PAH emission.}
\label{fig:mismatch}
\end{figure*}

In reducing the \sfmc\ observations, we found that the spectra were
affected by an artifact caused by a time- and positionally variable
background level on the detector.  This artifact occurred in both SL and
LL observations.  It appears to be similar to the ``dark settle''
artifact that can be important for high-resolution observations with
IRS.  The artifact causes mis-match in the regions where the orders
overlap and was first identified because it causes the long wavelength
end of SL2 and the short wavelength end of SL1 to be offset from each
other, sometimes significantly.  Since this wavelength region covers the
7.7 \micron\ PAH feature, understanding the effects of this artifact was
crucial for our study.  In Figure~\ref{fig:mismatch} we show several
spectra from the SW Bar 1 region illustrating the mismatches at SL and
LL caused by the artifact. Similar effects are seen in all maps, except
for the deep SMC B1 \#1 SL map, which used 60 second ramps and does not
appear to suffer from this artifact.

\begin{figure*}
\centering
\epsscale{0.9}
\plotone{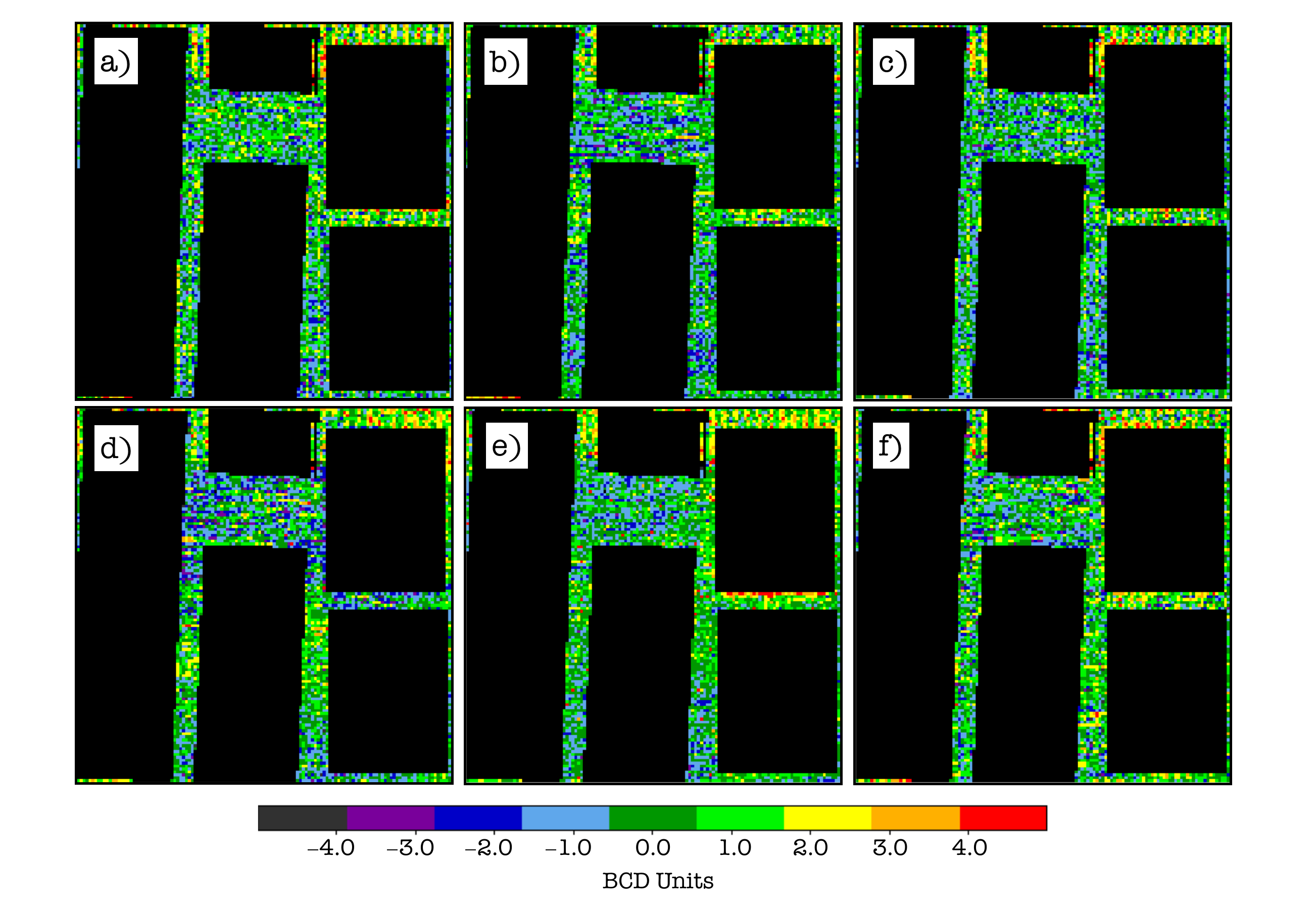}
\caption{The median of all background subtracted BCDs for the SL mapping
AORs in a) N22 North, b) N66 North, c) N 76 \#1, d) N 83, e) SW Bar 1
and f) SW Bar 3.  The SL1, SL2 and SL3 orders and the peak-up arrays
have been masked to more clearly show the background structure. These
images highlight the {\em non-time variable} characteristics of the
residual background, which are similar in each mapping AOR.  These
include a higher background level at the top of the detector ($\sim$8
\micron\ in the SL1 order and SL3) and in the middle of the detector (at
$\sim$ 5-6 \micron\ in the SL2 and $\sim$ 12 \micron\ in SL1).}
\label{fig:slmedbcd}
\end{figure*}

\begin{figure*}
\centering
\epsscale{0.9}
\plotone{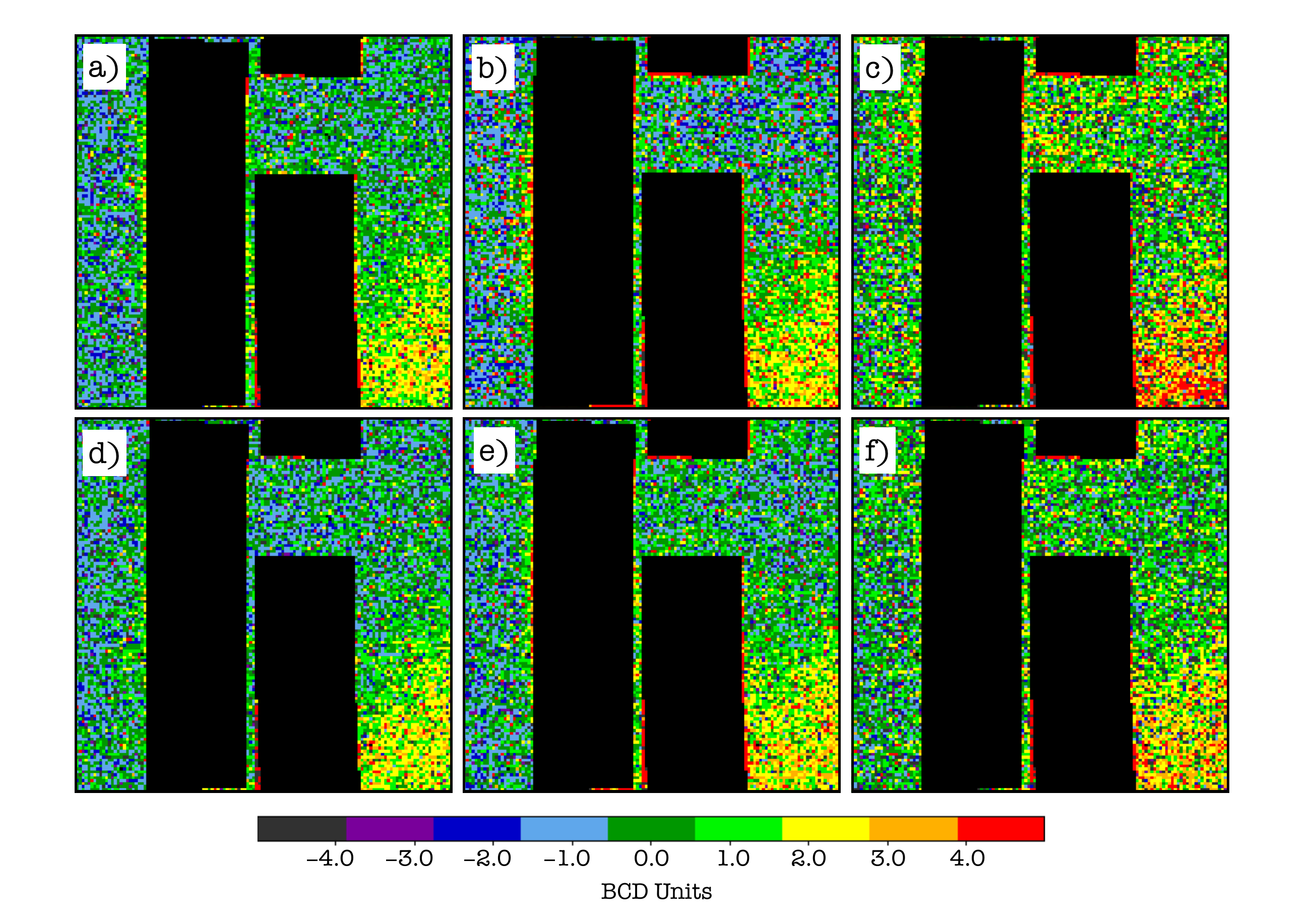}
\caption{The median of all background subtracted BCDs for the LL mapping
AORs in a) N22, b) N66, c) N76, d) N83, e) SW Bar~1 and f) SW Bar 3. The
LL1, LL2 and LL3 orders have been masked out to show the residual
background more clearly.  These images highlight the {\em non-time
variable} characteristics of the residual background.  For the LL maps,
the residual background is concentrated in the bottom right of the
detector.  This causes an offset at the long wavelength end of LL2,
which can be seen for SW Bar 1 in Figure~\ref{fig:mismatch} when the LL2
and LL1 orders do not agree in their overlap region. }
\label{fig:llmedbcd}
\end{figure*}

We found that the shape and magnitude of the offsets can generally be
explained by residual background emission that is apparent in the
inter-order region of the BCDs.  The residuals persists after removal of
the ``off'' observations and are much larger in magnitude than the
difference in zodiacal light or MW foreground between the map position
and the ``off'' position.  The average properties of this dark level can
be seen in the median background-subtracted BCD for each mapping
AOR---created by subtracting the median background image (created in
CUBISM) from each of the BCDs in a mapping AOR and then combining the
stack of all of the BCDs to generate one median background-subtracted
BCD.  Figure~\ref{fig:slmedbcd} shows these for each of our SL mapping
observations and Figure~\ref{fig:llmedbcd} shows these for each of our
LL mapping observations.  In these Figures we have masked out the orders
and the peak-up arrays to show the residual background more clearly.

\begin{figure*}
\centering
\epsscale{2.0}
\plottwo{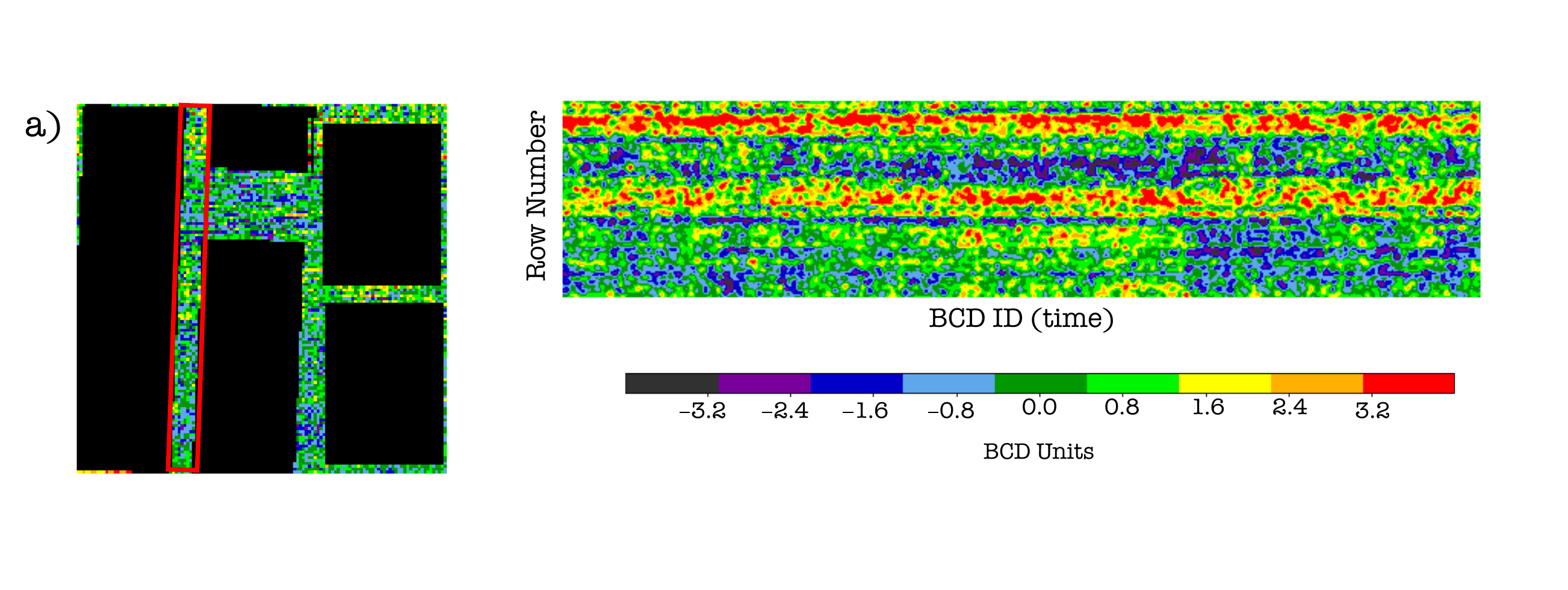}{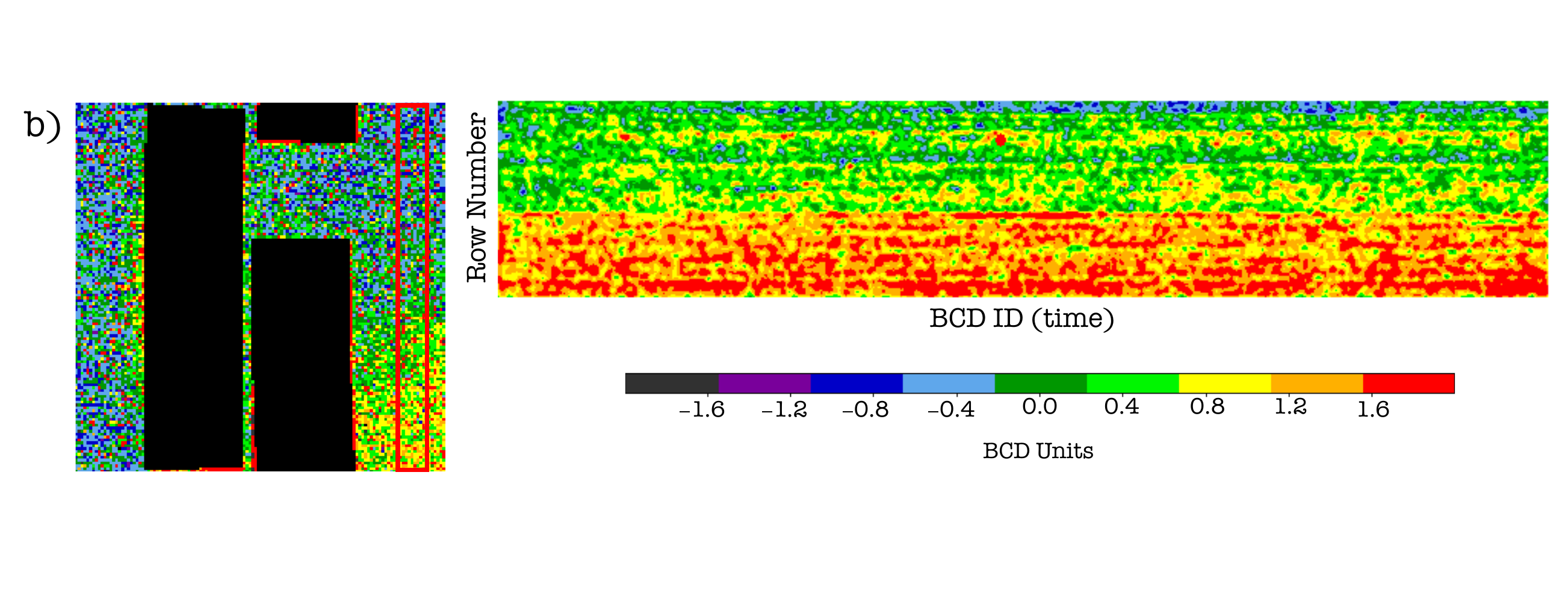}
\caption{The median value of the background subtracted BCD in a $\sim$5
pixel wide box (overlayed in red on the median BCD on the left) as a
function of BCD number, which generally corresponds to time in the AOR.
Panel a) shows the N66 South SL mapping AOR that includes 600 BCDs and
panel b) shows the N 66 LL mapping AOR that includes 686 BCDs.  The
images have been smoothed with a 3$\times$3 pixel Gaussian to suppress
noise. These images highlight the time-variable characteristics of the
residual background level.}
\label{fig:vstime}
\end{figure*}

In addition to the spatial structure of the background, there is also
some degree of time-variability.  In Figure~\ref{fig:vstime} we
illustrate the time variations of the residual background by displaying
the median of a $\sim$5 pixel wide strip of the background subtracted
BCDs versus BCD number (essentially a time sequence of the
observations).  

\subsection{Correction of the Artifact Derived from Inter-order Background Level}

To correct the effects of the artifact, we use the inter-order regions
to determine the spatial shape and time-variability of the residual
background.  This approach is very similar to that done by Starkey et
al. (2011, in prep). Because of pixel-based noise, we generally need to
average BCDs together as a function of time to improve our determination
of the background. At SL wavelengths the presence of the ``peak up''
arrays on the right-side of the BCD make the available inter-order
regions very small, since scattered light from the peak up arrays
affects the nearby inter-order regions.  We are therefore limited to the
small strip between the SL1 and SL2/SL3 orders to characterize the
vertical spatial dependence of the residual background (this region is
highlighted in Figure~\ref{fig:vstime}).  For the SL orders we determine
the median BCD value as a function of time and vertical row in a box
$\sim$5 pixels wide by 11 pixels high by 3 BCDs long.  We then create a
correction image for each BCD where all columns contains this median.
Finally, we apply the SL flatfield to get the appropriate value within
the orders.  

At LL wavelengths, the lack of peak up arrays make this process much
simpler.  In addition, there is far less time variability in the LL
residual background so we use a simpler correction derived from the
median BCDs shown in Figure~\ref{fig:llmedbcd}.  We mask the LL orders
and fit a horizontal polynomial to each row of the BCD, filling in each
row of the correction image with this fit.  We then apply a vertical
polynomial smoothing in a 10 pixel box and finally use the LL flatfield
image to obtain the correction in the LL orders.  For both the SL and LL
we subtract these correction images from the original BCDs.

The effect of the correction on several spectra can be seen in
Figure~\ref{fig:mismatch} in black.  The agreement between orders in the
overlap region is essentially always improved after the correction.
Most of this correction comes from removing flux in the 7$-$8 \micron\
region, coincident with the 7.7 PAH feature.  Because of the limited
information about the spatial structure of the residual background and
the large effect it has on the PAH feature, we use both the corrected
and uncorrected spectra to analyze the PAH band ratios in this paper.
Future work my reveal the source of this residual background and improve
our determination of the PAH feature strengths.

To conclude we make a few remarks about the sort of observations this
artifact is likely to affect.  Generally, the residual background level
is not an issue in typical IRS observations when either (1) an in-slit
background can be subtracted, as in the case of point source
observations; (2) the background observation is very close in time to
the mapping observation, for example when a ``nod'' position can be used
for background subtraction; or (3) when the object being mapped is
bright enough that these low-level variations are negligible.  In the
SMC we are observing faint emission using the full IRS slit and our
``off'' observation occurs after $\sim$600 pointings.  Thus, the \sfmc\
observations present a somewhat unique situation where the residual
background artifacts are important to remove.

{\it Facilities:} \facility{Spitzer ()}


\bibliographystyle{apj}

\end{document}